\pdfoutput=1

\documentclass[a4paper,10pt,twoside]{article}
\usepackage{pdfpages}
\usepackage{orcidlink}
\usepackage{xcolor}
\usepackage{authblk}
\newcommand{\rev}[1]{\textcolor{black}{#1}} 
\usepackage{placeins}   % \FloatBarrier
\usepackage{amsmath,amssymb}             % AMS Math
\usepackage[utf8]{inputenc}
\usepackage[T1]{fontenc}
\usepackage[left=1.5in,right=1.3in,top=1.1in,bottom=1.1in,includefoot,includehead,headheight=13.6pt]{geometry}

\usepackage{xcolor}
\usepackage{url,hyperref}
\usepackage{comment}
\usepackage{cite}
\usepackage{minitoc}
\usepackage{aecompl}
\usepackage{enumitem}

\usepackage[intoc]{nomencl}
\usepackage{booktabs}

\makenomenclature

\usepackage{color}
\definecolor{linkcol}{rgb}{0,0,0.4} 
\definecolor{citecol}{rgb}{0.5,0,0} 
\definecolor{jocol}{rgb}{0.1,0.6,0.3}

\usepackage{rotating}                    % Sideways of figures & tables
\usepackage{fancyhdr}                    % Fancy Header and Footer

\let\headruleORIG\headrule
\renewcommand{\headrule}{\color{black} \headruleORIG}

\usepackage{colortbl}
\arrayrulecolor{black}

\fancypagestyle{plain}{
  \fancyhead{}
  \fancyfoot{}
  
}

\usepackage{algorithm}
\usepackage[noend]{algorithmic}

\makeatletter

\def\cleardoublepage{\clearpage\if@twoside \ifodd\c@page\else%
  \hbox{}%
  \thispagestyle{empty}%              % Empty header styles
  \newpage%
  \if@twocolumn\hbox{}\newpage\fi\fi\fi}

\makeatother
 
{%

\hrulefill
\vspace*{0.5cm}%
\end{minipage}
}

\usepackage{multirow}
\usepackage{slashbox}
{ \begin{list}%
	{$\bullet$}%
	{\setlength{\labelwidth}{25pt}%
	 \setlength{\leftmargin}{30pt}%
	 \setlength{\itemsep}{\parsep}}}%
{ \end{list} }

\renewcommand{\epsilon}{\varepsilon}
\usepackage{enumitem}
\newlist{myitemize}{itemize}{2}
\setlist[myitemize,1]{label={},leftmargin=0em}
\setlist[myitemize,2]{label=o,leftmargin=1em}
\newlist{youitemize}{itemize}{2}
\setlist[youitemize,1]{label=$\blacktriangleright$,leftmargin=*}
\setlist[youitemize,2]{label=o,leftmargin=1em}
\usepackage{upgreek}
\usepackage{textgreek}

\newcommand{\np}{n+p}
\newcommand{\pp}{p+p}
\newcommand{\pim}{$\uppi^{-}$}
\newcommand{\pip}{$\uppi^{+}$}
\newcommand{\piz}{$\uppi^{0}$}
\newcommand{\pimP}{$\uppi^{-}$+PE}
\newcommand{\pimp}{$\uppi^{-}$+p}
\newcommand{\pimC}{$\uppi^{-}$+C}

\newcommand{\pimAg}{$\uppi^{-}$+Ag}

\newcommand{\gevcc}{GeV/$c^{2}$}
\newcommand{\mevcc}{MeV/$c^{2}$}
\newcommand{\gevc}{GeV/$c$}

\newcommand{\gev}{GeV}
\newcommand{\mev}{MeV}

\newcommand{\sqrts}{$\sqrt{s}$}

\newcommand{\ee}{e$^+$e$^-$}
\newcommand{\mee}{$M_\text{ee}$}

\newcommand{\lh}{LH$_2$}

\begin{document}

\DeclareGraphicsExtensions{.jpg,.gif,.pdf,.png,.mps,.eps}

\includepdf[pages=-]{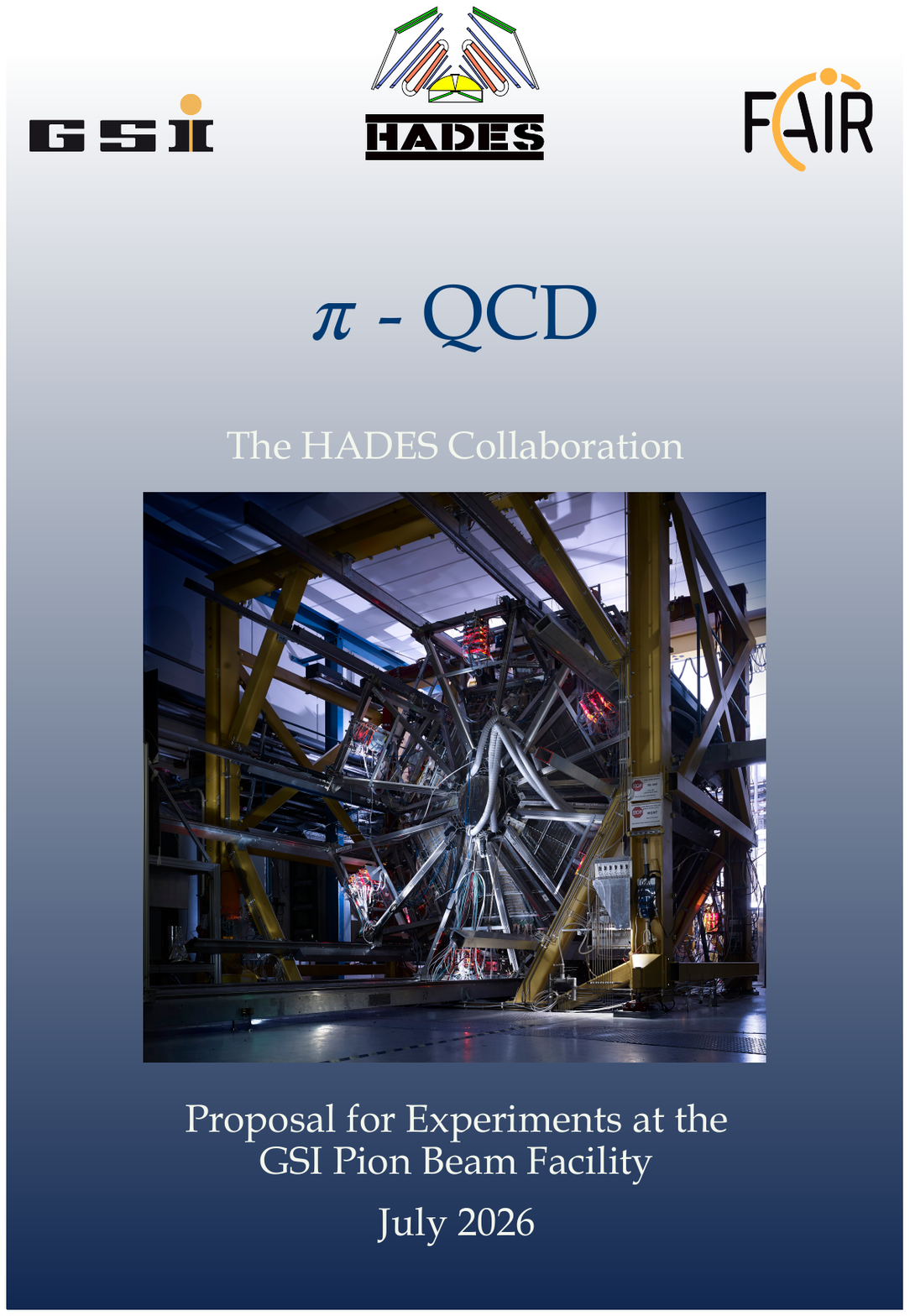}
\newpage
{\em \bf Abstract}

  We propose to investigate the dynamics of hadronic matter in the strong QCD regime using the pion-beam facility at GSI in conjunction with HADES. This program will study hadron physics through elementary pion-nucleon interactions up to $\sqrt{s}\approx$2.35 GeV, bridging cold matter research with detailed analyses of exclusive reactions.
 Our goals include exploring baryon resonance formation and their electromagnetic structures. This will complement photo-production experiments (e.g., ELSA) and enhance our understanding of emissivity in dense, hot hadronic matter. Key topics include examining baryon couplings to mesons and virtual photons, analyzing hyperon weak decays in exclusive reactions, and using Partial Wave Analyses with differential cross sections and polarization observables to achieve unprecedented precision in baryon-meson coupling data ($\uprho$N, $\upomega$N).
Furthermore, \ee\ production measurements off nucleons will provide insight into the electromagnetic transition form factors of baryons in the time-like region, revealing the role of vector mesons ($\uprho$, $\upomega$). 
Reactions on nuclear targets allow for detailed studies of hadron properties, including vector-meson line shapes and strengths in cold nuclear matter. \rev{These measurements provide a critical reference for the interpretation of results from the hot and dense environment created in $A+A$ collisions.}
This research also has broader implications. It will contribute to the modeling of neutrino-nucleus interactions, facilitate unprecedented investigations into hypernuclei formation, and enhance our understanding of strong interaction dynamics across different energy regimes.

\newpage
\tableofcontents
\newpage
\section{\rev{Executive summary}}
%-------------------------------------------------------------
\label{sec:state-art}

\subsubsection*{Pion-matter interactions as tool for a collaborative QCD program}

  Pion-nucleon scattering has been fundamental to understanding nuclear binding forces, with pions serving as the (pseudo) Goldstone bosons of SU(2) flavor symmetry. The $\pi$N system is a well-established probe of strong QCD dynamics, attracting significant
  experimental and theoretical interest. Experimentally, pion-induced processes are relatively straightforward to interpret compared to, for example, nucleon-nucleon scattering. With a baryon number of one, these interactions typically produce two or three particles in the final state, facilitating rigorous partial-wave analyses (PWA) and allowing for high detection efficiencies that support the reconstruction of complete event topologies. Compared to complementary photo-production experiments, $\pi$N cross sections are substantially larger, enabling high-precision measurements within shorter beam times, making them an efficient tool for advancing our understanding of hadron dynamics.

  \rev{This document defines the physics programme for future experiments with pion beams at GSI/FAIR using HADES and provides the scientific framework for subsequent, experiment-specific beam-time proposals. It is not meant as a closed programme, but can evolve over the coming years as new physics opportunities and experimental possibilities emerge.}

\begin{figure}[tbh]
  \begin{center}
    \includegraphics[width=0.48\textwidth]{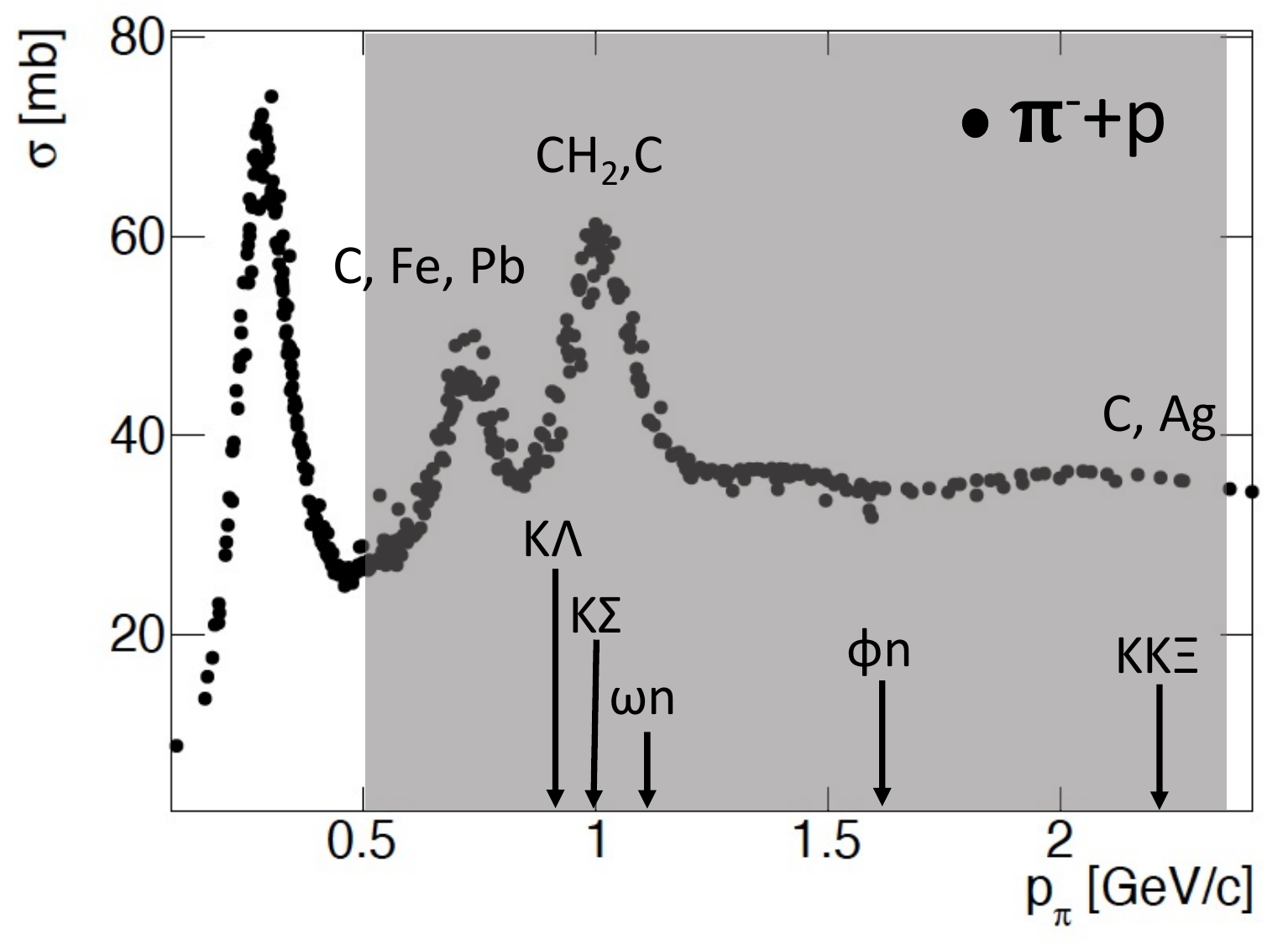} 
    \includegraphics[width=0.5\textwidth]{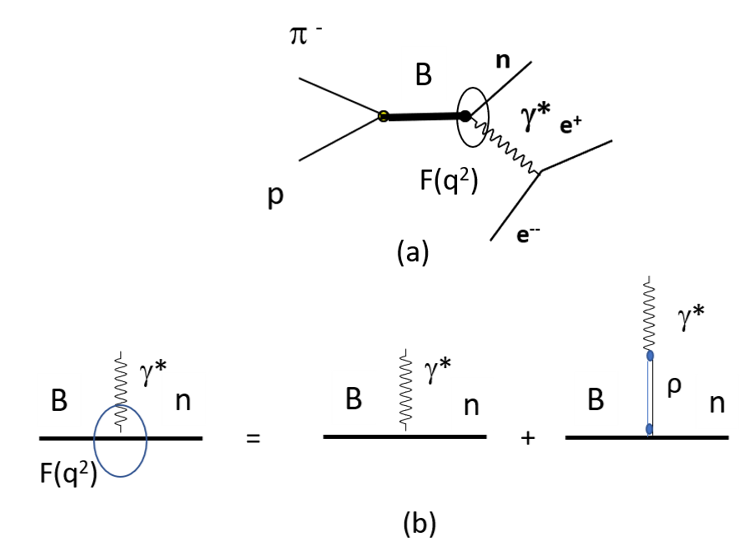}  
  \end{center}
   \scriptsize
  \caption{\scriptsize Left: \rev{Compilation} of \pimp\ cross sections (black dots) \cite{ParticleDataGroup:2026aaa}. The black arrows indicate the thresholds for various hadronic final states. The gray shaded area corresponds to the momentum range covered by this proposal, the targets in the upper part mark the regions, in which they are used.  
  Right: (a) Sketch of a time-like electromagnetic baryon transition in the \pimp\ reaction at a given four-momentum transfer q$^2=$\mee $^2$. 
  (b) illustration of the two-component form factor model, consisting of a point-like photon coupling and an electromagnetic interaction mediated by vector mesons, according to the Vector Meson Dominance (VMD) model.}
  \label{fig:VMD}
\end{figure}

  A common interest in the nuclear, hadron, and heavy-ion physics communities has evolved towards understanding strong QCD guided by ``SU(3) flavored systems'', addressing, {\it f.e.}, the role of strangeness in dense baryonic matter, such as expected to occur in the core of neutron stars. The	program	proposed in this document has the ambition to provide valuable data in this context by exploiting pion-beam experiment on proton and nuclear targets in the pion momentum 0.5-2.5 \gev/c\, see Fig.\ref{fig:VMD}, Left (a). The program makes use of the unique combination of an intense pion beam (10$^6$/s) and the versatile HADES setup, which provides extensive phase-space coverage, precise momentum determination, state-of-the-art di-lepton detection, excellent particle identification, and photon detection capabilities via its newly completed electromagnetic calorimeter. These features enable precision measurements of a wide range of observables, including differential cross sections, branching fractions, time-like electromagnetic transition form factors, spin-density matrix elements, and the self-polarization of hyperons, for various reaction channels below and above the hyperon-production threshold.

    In studies of cold matter, the use of the $\pi$A initial state offers several key advantages \rev{compared to $p$A collisions}. Most prominently, its kinematics allows to produce short-lived hadrons, such as the $\omega$ vector meson, with relatively small momenta \rev{with respect to the recoil nucleus}. This provides promising sensitivity for investigating in-medium effects, e.g. line shape and line strength measurements of vector-mesons, which are expected to manifest at low momenta. Secondly, pion-induced reactions are superior to the previously studied proton- and photon-induced reactions. Due to the large $\pi$N inelastic cross section, hadron production occurs close to the upstream surface of the nucleus, leading on average to a longer path of the produced hadrons inside the nuclear matter. Additionally, the high production cross-section of hyperons in $\pi$-induced reactions makes it particularly well-suited for the abundant formation of hypernuclei. Pion-induced cold matter studies also yield data that are crucial for testing and refining transport models. The significance of such datasets has been highlighted by neutrino experiments, such as T2K and DUNE, which \rev{depend on accurate hadron-production and interaction models for the reconstruction of the incident neutrino energy}.

\subsubsection*{Pion beams at GSI: a proven success story}

\begin{figure}[t]
  \begin{center}
          \includegraphics[width=1.0\textwidth]{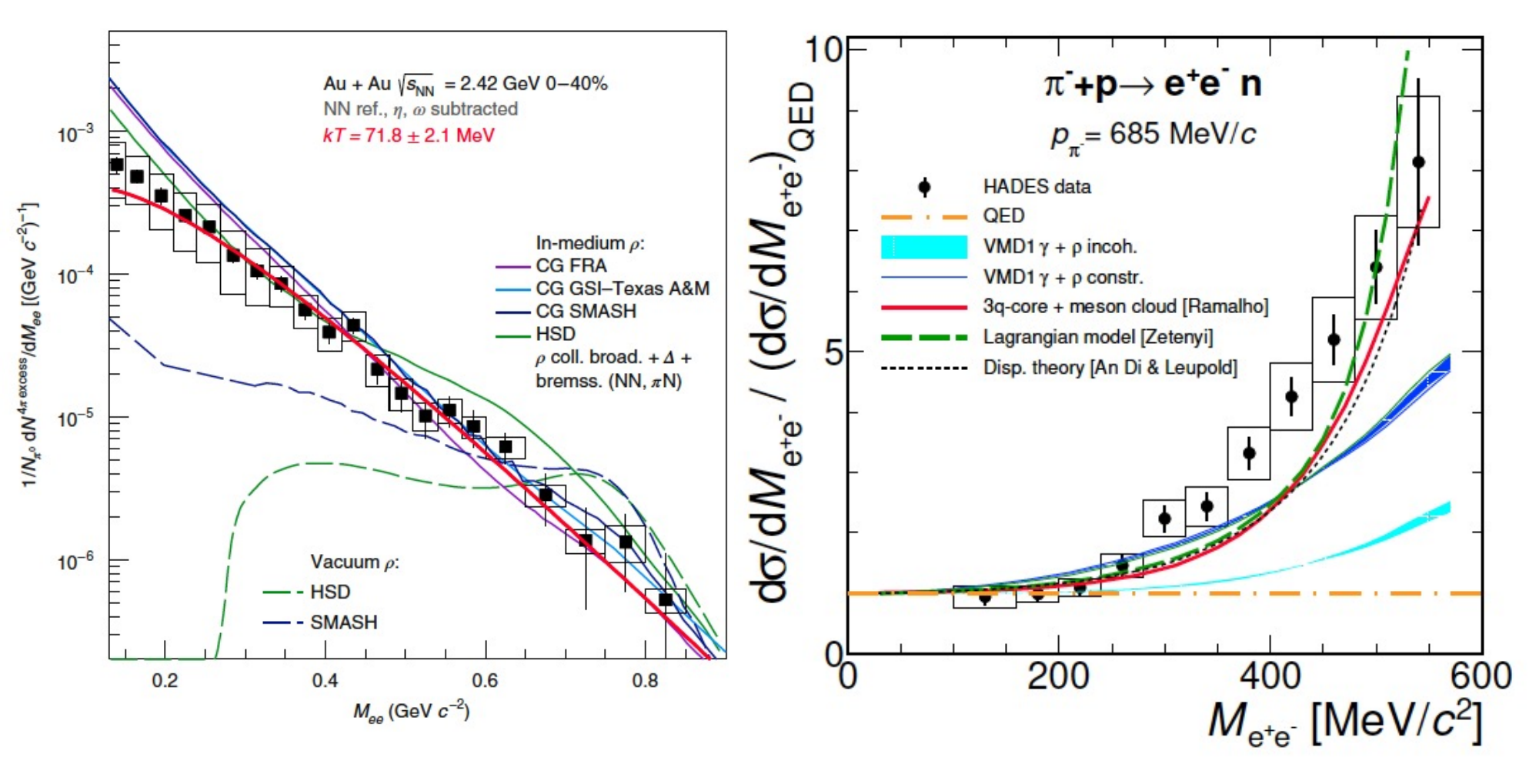}   
          \vspace*{-5mm}
        \caption{\scriptsize Left: Dilepton excess yield in the low invariant mass region identified as the modified $\rho$-spectral function. Different theoretical calculations are compared to the data measured by HADES in Au+Au collisions at 2.42 GeV~\cite{Adamczewski-Musch:2019byl}. Right: Invariant mass of pairs from $N^*\rightarrow N\gamma^*\rightarrow N e^+e^-$ normalized by QED prediction for point-like transitions~\cite{AbouYassine:2022hmt}.}
        \label{fig:NuPECC_LRP2024}
  \end{center}
\end{figure}

 The physics potential of pion-induced reactions combined with HADES, particularly for \rev{connecting hadron and heavy-ion physics}, has been unequivocally demonstrated in previous studies. A notable example is the investigation of the $N^*(1520)\rightarrow N e^+e^-$ electromagnetic transition form factor (emTFF) in $\pi^- p \rightarrow N e^+e^-$ at a center-of-mass energy of $\sqrt{s} = 1.5~\text{GeV}$. The $q^2$ dependence of the emTFF, together with a partial-wave analysis of simultaneously reconstructed $\pi^-p \rightarrow \pi^+\pi^-n$ data, provided detailed insights into the validity of vector-meson dominance and the role of pion clouds. This information is crucial for understanding $e^+e^-$ mass spectra in $A+A$ reactions, especially in the context of the $\rho$ contribution melting in baryon-dense environments as illustrated in Fig.~\ref{fig:NuPECC_LRP2024}.

  The proposed program exploiting the pion-beam and HADES facilities enables detailed investigations of both hadronic and electromagnetic properties of matter. The versatility and efficient data acquisition capabilities of the combined facility make it a powerful tool for connecting diverse physics topics across multiple fields. In this document, we outline various topics, intimately connected, that we wish to follow-up in the upcoming years.

\subsubsection*{\rev{A QCD-driven roadmap for experiments at GSI/FAIR}}

\begin{figure}[t]
  \begin{center}
          \includegraphics[width=1.0\textwidth]{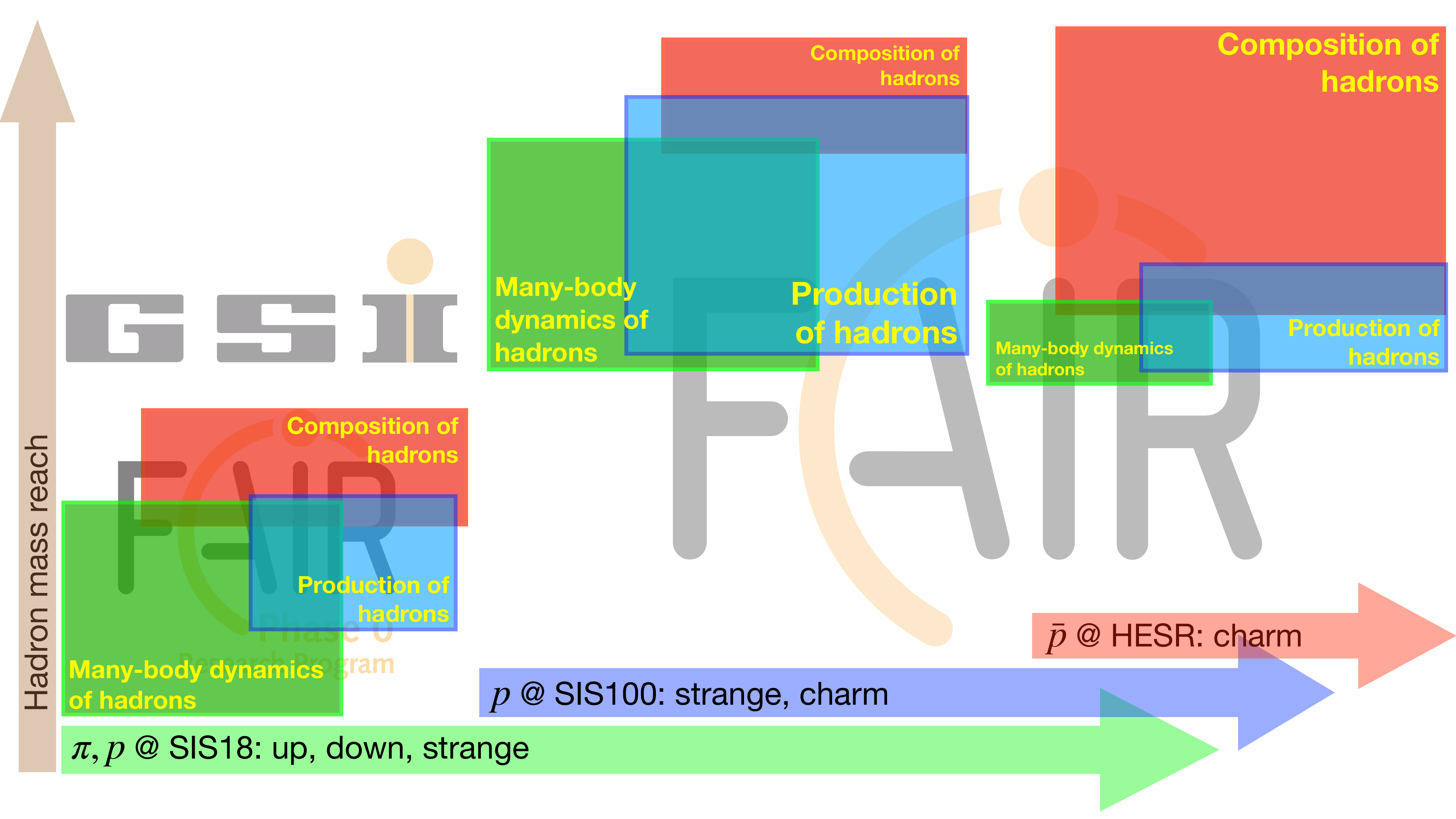}   
          \vspace*{-5mm}
        \caption{\scriptsize An illustrative sketch of the roadmap for hadron physics activities at GSI and FAIR. \rev{The green arrow indicates the pion- and proton-beam programme at SIS18, while the blue and red arrows indicate the proton programme at SIS100 and the antiproton programme at HESR, respectively, with increasing hadron-mass reach.} The three categories of boxes, distinguished by green, blue, and red, symbolize the three research domains to be tackled. The size of each box reflects its respective contribution to the overarching program.}
        \label{fig:roadmap}
  \end{center}
\end{figure}

The physics program exploiting the pion-beam facility in combination with HADES is part of a broader hadron-physics motivated program \rev{accompanying the transition from GSI to FAIR}. The combination of hadronic beams provided by SIS100 and versatile detector systems, such as CBM, \rev{will allow us to pursue a long-term and QCD-motivated program} involving the three communities (nuclear, hadron, and heavy-ion physics) and driven by the experiences as obtained with the pion-beam facility and HADES. \rev{SIS100 enables us to reach center-of-mass energies in the proton-proton system of about $\sqrt{s}$=7.6~GeV, thereby allowing us to probe baryonic systems with single, double, and even triple strangeness, as well as charm-enriched matter.} With unprecedented beam intensities, combined with high-acceptance and versatile detector systems capable of handling high-interaction rates and utilizing free-streaming data processing, FAIR will enable precision studies in the field of QCD. \rev{As part of the further development of FAIR, the modularized start version (MSVc) foresees the implementation of the high-energy storage ring (HESR),} specially designed to contain beams of antiprotons within an energy spectrum ranging from 0.8~GeV to 14~GeV, stochastically cooled to achieve a momentum spread of approximately $\Delta p/p\approx 10^{-5}$. \rev{Combined with the versatile 4$\pi$ PANDA detector, the HESR will enable a wealth of complementary hadron-physics studies, in particular precision measurements in hadron spectroscopy and structure}~\cite{PANDA:2021ozp}. Figure~\ref{fig:roadmap} illustrates the foreseen roadmap at GSI/FAIR.

\subsubsection*{Structure of the document}

This document is organized according to the three themes that we aim to address with the pion-beam facility. In Sec.~\ref{sec:matter} we highlight some of the topics that can be explored in the field of cold matter with nuclear targets. Section~\ref{sec:structure} addresses how pion-nucleon interactions can be used to study the composition of hadronic matter. Subsequentially, in Sec.~\ref{sec:effinteractions} we describe the physics associated with the interactions among hadrons. In Sec.~\ref{sec:exp-setup}, we provide a concise description of the experimental capabilities related to the pion-beam facility and HADES. 
\clearpage
\section{{\it Q:} Cold matter}
\label{sec:matter}

\begin{figure}[tbh]
  \begin{center}
    \includegraphics[width=0.68\textwidth]{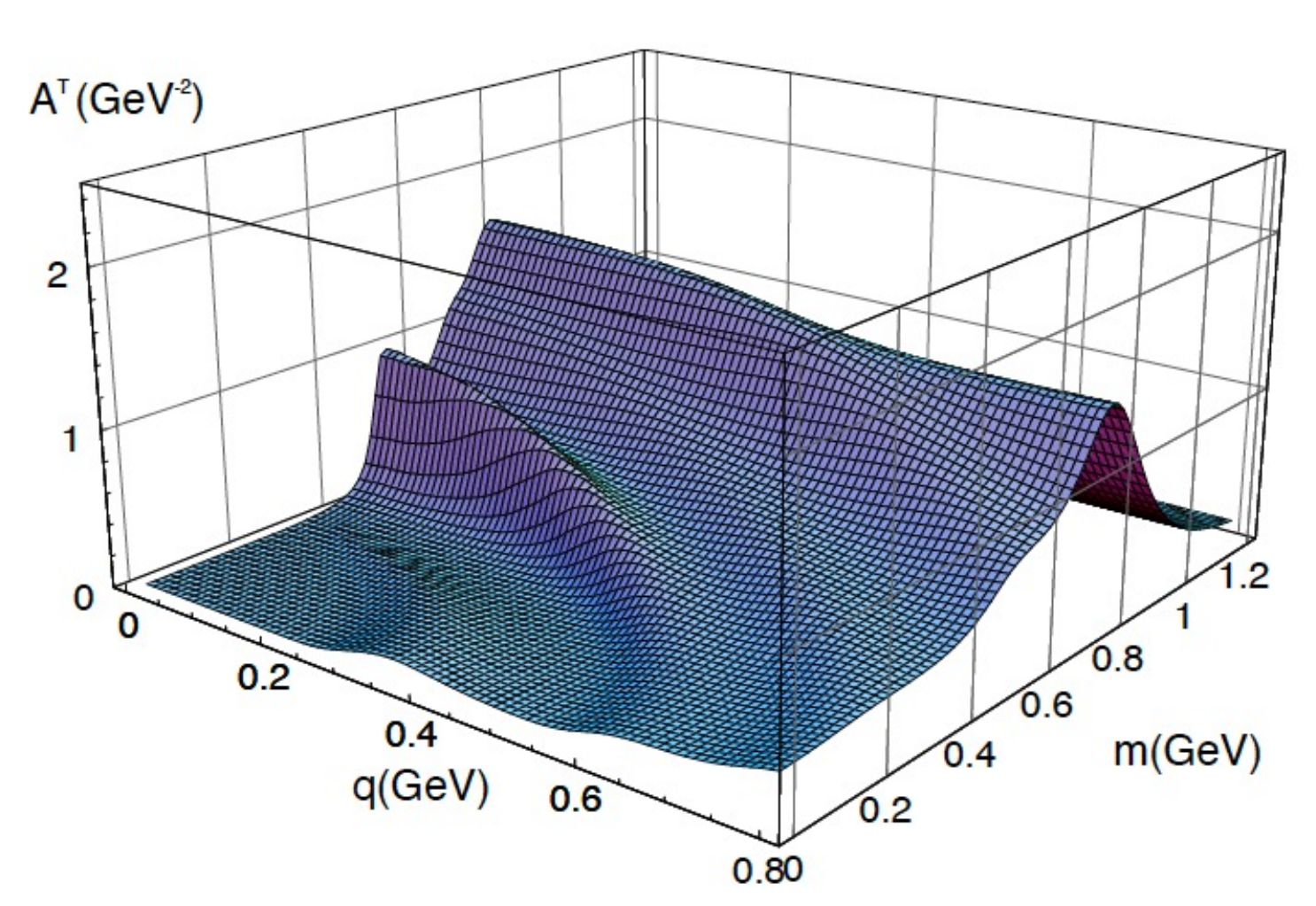} 
  \end{center}
   \scriptsize
   \vspace*{-0.5cm}
  \caption{\scriptsize Transverse spectral function of the $\rho$ meson at twice the nuclear saturation density, calculated using a hadronic model~\rev{\cite{PostMosel2002}}. The most significant modifications are observed at small relative momenta, highlighting the medium's pronounced influence in this regime.}
  \label{fig:rhospec}
\end{figure}

\subsection{Introduction}

A key focus of HADES, an exceptional dilepton spectrometer, is to investigate the properties and baryon couplings of vector mesons in the nuclear medium. The use of pion beams in these studies offers notable advantages, particularly due to the ideal kinematical properties of the final-state particles, such as their small relative momenta, which enhance sensitivity to medium effects. In addition to exploring the propagation of vector mesons in cold matter, it has become clear that the highest momentum pion beams present a unique opportunity to study the \rev{production of particles with hidden- and open-strangeness in the nuclear medium}. Consequently, the optimal energy for these cold nuclear matter studies is $\sqrt{s}=2.35~\text{GeV}$, which exceeds the production thresholds for $\phi$-mesons and $\Xi$-hyperons. Overall, the total strangeness yield will increase, promoting the formation of kaons, single-strange hyperons, and hypernuclei. To achieve this energy with sufficient intensity, a high-energy primary proton beam is essential, ideally at the maximum available energy.

In addition to focusing on the high-energy part of the spectrum, there is growing interest in collecting precision data with pion beams to provide reference measurements for modeling neutrino-nucleus reactions. This has attracted significant attention from the neutrino research community. Such an experimental program can largely be carried out at lower energies, for example, at $\sqrt{s}=1.76~\text{GeV}$ using a $^{14}$N primary beam, albeit at the cost of studying $\phi$-mesons and $\Xi$-hyperons.

In the following, we briefly outline the most promising cold matter topics we identified that can be addressed at the pion-beam facility in combination with HADES. \rev{The hadron spectroscopy and hypernuclei aspects will be addressed in Sections~\ref{sec:structure} and~\ref{sec:effinteractions}, respectively.}\\

\subsection{In-medium vector-meson properties}

Combining dilepton spectroscopy with a $\pi$-beam offers a unique set of advantages for studying in-medium hadron properties. When compared to a p-beam, a $\pi$-beam brings in less momentum, resulting in secondary particles being produced with small recoil momentum. This recoil-less kinematics allows the particles to interact with the nuclear medium for a longer time. For penetrating probes such as the decay of vector-mesons reconstructed via their dielectron decays, this setup significantly enhances the inside-to-outside fraction. The HADES experiment has a unique advantage over other experiments such as CLAS or KEK \cite{clas,kek}. It offers extensive coverage of pairs with low momenta relative to the nuclear medium, enhancing the expected modification of hadrons in nuclear matter. This is illustrated in Fig.~\ref{fig:rhospec}, which presents the transverse spectral function of the $\rho$ meson at twice the nuclear saturation density, calculated using a hadronic model~\cite{post}. \rev{The calculation at twice saturation density is shown here for illustrative purposes and is not meant to represent the density reached in pion-induced reactions; it makes particularly transparent that the largest medium modifications occur at low relative momentum.} The spectral function is shown as a function of the relative momentum with respect to the medium. The most significant modifications are observed at small relative momenta, highlighting the medium's pronounced influence in this regime.
This makes HADES combined with a pion beam the ideal place to investigate line shape modification and line strength suppression due to in-medium properties of the light vector mesons $\rho, \omega,$ and $\phi$ as a function of momentum.\par
A breakthrough in cold matter studies was achieved by the measurements with HADES of the p+Nb reaction at 3.5 GeV, thanks to the capability to measure dielectrons close to target rapidity, \textit{i.e.}, at low relative momentum to the cold matter \cite{Agakishiev12_pNb}. These measurements have shown a substantial modification of the in-medium $\uprho$ meson and a strong absorption of the $\upomega$. Thanks to the favorable kinematics and the specificity of pion-induced reactions to excite baryon resonances in a given range, we expect a high scientific impact from measurements in the $\pi^-+$Ag reaction.
The measurements on the carbon target will provide a reference. \rev{Some nuclear effects, such as Fermi motion, are already present in the carbon reference.} The silver target, however, will enhance the number of decays inside the nucleus and allow investigation of effects related to changes in hadron properties in cold nuclear matter.
Our primary motivation is to measure medium effects that modify the $\uprho$ and $\upomega$ vector meson properties, as predicted in \cite{Effenberger99nn}. We estimated count rates for the $e^+e^-$ yields from $\uprho$ and $\upomega$ mesons produced in the $\pi^-+$Ag reaction, based on a quasi-free process and scaling with $Z^{2/3}$ (see Table~\ref{tab:ratesPEC} in Sec.~\ref{sec:structure}). This scaling provides an estimate of the expected yields in the absence of medium effects, which are likely to produce an excess of dielectrons below the vector meson pole due to the modified $\uprho$ contribution. 
A reduction of the line strength in the $\upomega$ pole region in data from the Ag target, with respect to data from the C target, is also a sensitive observable for broadening of the hadron in the medium, which will be addressed. If data is taken at $\sqrt{s}=2.35~\text{GeV}$, the $\phi$ meson can be investigated within the same experimental run.\\
\\
\subsection{Production and propagation of strangeness}
Strangeness-carrying hadrons serve as a sensitive and penetrating probe for the properties of the medium. Additionally, the kaon-nucleon and hyperon-nucleon interactions are crucial for many astrophysical processes, such as our understanding of the stability of neutron stars. The excellent particle identification (PID) capabilities and momentum resolution of HADES enable the simultaneous measurement of charged kaons, $\Lambda$ hyperons, and $\phi$ mesons (via their charged kaon decay channels). This facilitates a detailed study of the production and propagation of strangeness in cold nuclear matter, complementing the information gained from dilepton studies and providing relevant inputs for astrophysical questions, including neutron star stability.
In particular, kaons will be used to study the kaon-nucleus potential. This has been demonstrated with HADES data from p+Nb reactions at 3.5~GeV~\cite{Agakishiev12_pNb}, and has also been investigated in the most recent $\pi^-+$W and $\pi^-+$C reactions at $\sqrt{s}=2.0~\text{GeV}$~\cite{HadesStrangenessPionbeam}, where strong absorption of $K^-$ and $\phi$ mesons was observed. Furthermore, the formation of $\Lambda$ and $\Xi$ hyperons will be addressed. In particular, the production rate of $\Xi$ hyperons in cold nuclear matter is crucial for understanding the observed enhancement in heavy-ion collisions.
In general, such data remain scarce and, as mentioned above, are essential for constraining the equation of State~\cite{HadesStrangenessPionbeam}.\\
\\
\subsection{Intermediate reference for heavy-ion data}
Pions are abundantly produced in heavy-ion collisions and have a high probability of interacting with the nuclear medium. 
Thus, pion-nucleus reactions are essential for refining theoretical models of medium effects in hot and dense nuclear matter. 
Pion-induced reactions offer the advantage of selectively exciting a narrow mass range of the baryon-resonance spectrum, unlike proton-nucleus reactions, which excite a broader spectrum dominated by the $\Delta(1232)$ resonance, especially at beam energies up to 3~GeV. Pion-beam interactions provide essential information about resonances with pole masses above that of the $\Delta(1232)$, data that is currently lacking for higher-lying resonances. This information will be invaluable for experiments at higher energies, such as those planned at FAIR.
In summary, pion-nucleus reactions represent a critical intermediate step between elementary nucleon-nucleon or pion-nucleon collisions and the hot and dense matter created in heavy-ion (A+A) collisions. Mesons (pions, $\upeta$, kaons, ...) as well as protons and light nuclei will be copiously produced, and their production spectra will serve as valuable benchmarks for hadronic models.\\
\\
\subsection{Input for neutrino-nucleus reaction modeling}\label{secneutrino}
The Long Baseline (LBL) Neutrino Oscillation Experiments are entering a new era of precision physics. \rev{In these experiments, neutrinos are produced by high-intensity accelerators (e.g., J-PARC, Fermilab).} The probability of neutrino oscillations depends on neutrino energy, mixing angles, and masses, and is precisely measured by comparing the neutrino rates, energies, and flavors at near detectors (close to the source) and far detectors (located hundreds of kilometers away).\par
Accurate measurement of the oscillation probability directly depends on neutrino energy reconstruction. The next generation of experiments, e.g. upgraded T2K, which began taking data this year, and the future Hyper-K and DUNE (operational by 2028/2029), will deliver large statistics~\rev{\cite{HyperKDesign,DUNETDR}}. However, reducing systematic uncertainties from the current $\sim 5-10\%$ to about 1-2\% is crucial for achieving their physics goals~\rev{\cite{NuSTEC2018,TENSIONS2016}}. The largest and most challenging uncertainties arise from modeling neutrino-nucleus interactions, primarily using hadronic models implemented in GEANT4. A longstanding issue in the LBL community is the discrepancy between models and neutrino cross-section measurements, especially in final states involving pion production. In T2K/Hyper-K, neutrinos in the 500-700 MeV energy range are detected via charged-current quasi-elastic interactions ($\nu_{\mu} + $p$ \to \mu^{-} + $n) with nuclei as targets. The neutrino energy is inferred from the measured kinematics of the outgoing lepton. However, single-pion production becomes significant in the high-energy tail of the neutrino flux, particularly above 1~GeV.
For DUNE, where the neutrino flux will be shifted to higher energies, pion production channels will play an even more significant role. Neutrino energy reconstruction will rely on detecting not only leptons but also pions and nucleons ($\nu_{\mu} + $p$ \to \mu^{-} + $n$ + N\pi$, where $N \geq 1$).\par
In this context, nuclear effects are critical, both at the interaction vertex and through re-scattering of produced pions within the nucleus, known as Final State Interactions (FSI). Other effects, such as Fermi motion and short-range correlations must also be accurately \rev{modelled}, as well as secondary interactions within detector materials.\par
Precise hadronic models must address these factors as well as contributions from missing energy, such as undetected neutrons (relevant for T2K), re-absorbed pions, protons, or energy lost below detection thresholds (low-energy hadrons). Reliable simulations are required to correct for such systematic effects.
Pion beam data are invaluable for improving model descriptions of neutrino-nucleus interactions. \rev{While the primary interaction differs between pion- and neutrino-induced reactions, the subsequent propagation and reinteraction of hadrons in the nucleus are governed by the same nuclear properties and processes, including Fermi motion, mean-field potentials, baryon-resonance excitation, pion rescattering, and absorption.}
However, the available pion beam data is scarce and largely limited to the $\Delta(1232)$ resonance region ($p \leq 500$ MeV/$c$). Crucially, there is a lack of data in the energy range essential for neutrino physics, particularly for interactions like $\pi^{-} + \text{Fe}$, $\pi^{-} + \text{Pb}$ (important for T2K), and $\pi^{-} + \text{Ar}$ (important for DUNE), which can be filled with the current proposal.\par
In the previous pion beam experiment in 2014 we have studied inclusive production of p,~$\pi^{+}$,~$\pi^{-}$,~d,~t  and semi-exclusive 2- and 3-particle (2$\pi^{-}$, 2$\pi^{+}$, 2p, $\pi^{+}$$\pi^{-}$p...) coincident spectra with the carbon target at 0.69 GeV/$c$ \cite{hojei23,ram24}.  \rev{The measurement also provided high-statistics data at four additional pion-beam momenta of 0.61, 0.66, 0.75, and 0.8 GeV/$c$.} \rev{The data at 0.69 GeV/$c$ have been used to benchmark the transport models SMASH} \cite{Petersen:2018jag}, \rev{RQMD.RMF} \cite{Nara:2021fuu}, \rev{GiBUU} \cite{Buss:2011mx} \rev{and the intranuclear cascade model INCL++, which are employed by the heavy-ion and neutrino-physics communities} \cite{PhysRevC.90.054602}. \rev{The transport-model predictions show an unexpectedly large dispersion among each other and with respect to the data, which points to the need for further adjustments.} 
\rev{In this proposal we plan to complement the existing high-statistics $\pi^{-}$+C data with a measurement at 0.5 GeV/$c$ and extend the database with measurements at 0.5, 0.6 and 0.7 GeV/$c$ using Fe and Pb targets. The selected energies and targets are relevant for constraining pion--nucleus interactions in the energy range of interest for long-baseline neutrino experiments.} The results will allow to test selectively the capacity of the models to describe various mechanisms like quasi-elastic, multi-pion production, rescatterings, and pion absorption processes. 
\\

\clearpage
\section{{\it C:} Hadron spectroscopy, structure, and exotics}
\label{sec:structure}

\subsection{Introduction}

The results extracted from the 2014 pion-beam data on  proton target collected within the second baryon resonance regime at $\sqrt{s}=1.5$~GeV\rev{~\cite{Hades17_pibeam}}, highlighted the benefits and unique features of utilizing the $\pi N$ interaction for baryon spectroscopy and structure studies. The $N^*$ selectivity achieved in the $\pi N$ scattering process via the s-channel offers a significant advantage over proton beams. The well-defined initial state, combined with the limited number of final-state particles, facilitates partial wave analysis, enabling the extraction of various meson couplings with excited baryons. Of particular interest are the $N^*\rightarrow \rho N$ couplings, which are highly relevant for the complementary heavy-ion program conducted with HADES. Beyond the detailed study of hadronic aspects in $\pi N$ interactions, the 2014 data demonstrated the unique capability to extract valuable electromagnetic structure information by measuring the timelike form factors in the $N^*\rightarrow N e^+e^-$ process, underlining the exceptional dilepton capabilities of HADES \rev{(see Fig.~\ref{fig:PionBeam2014})}.

\rev{Pion--proton reactions are accessed experimentally using polyethylene targets, with dedicated carbon-target measurements used to determine and subtract the contribution from pion--carbon interactions. This approach can be applied to both $\pi^-p$ and $\pi^+p$ reactions.}

The remarkable results pave the way for the planned pion-beam experiments discussed in this paper. Of particular interest is the exploration of the third resonance regime around $\sqrt{s}=$1.7~GeV, employing the same proven methodology. Furthermore, at these higher energies, the couplings of baryon resonances excited in $\pi N$ interactions to final states involving strangeness will become accessible. This will enable a systematic investigation of QCD within the framework of SU(3)-flavor symmetry. These anticipated $\pi N$ studies will complement planned photoproduction experiments at CB-ELSA.  

In addition to advancing studies in the baryon spectroscopy and structure sector, it will also be possible to extract valuable information on (exotic) meson-like states and glueballs through partial-wave analysis of $\pi\bar{\pi}$ and $K\bar{K}$ final states produced in $\pi N$ interactions at various energies. Furthermore, pion-nucleon scattering offers an ideal signal-to-background environment for investigating rare $\eta$ decays.

\subsection{Baryon-meson couplings}

\subsubsection*{Motivation}
Together with photon-nucleon reactions, pion-nucleon reactions provide direct information on the electromagnetic and hadronic couplings of baryonic resonances, which are related to their intrinsic structure in terms of quarks and gluons through hadron-structure models. While the database for photon-induced reactions has been continuously developed in recent years, \rev{progress in determining the baryon spectrum and related decay properties} is currently limited by the lack of precise data from pion beam experiments \cite{Briscoe15}. 

A better determination of the couplings of baryonic resonances to $\uprho$N and $\upomega$N final states is particularly important for the understanding of the \rev{dilepton emissivity of hot and dense nuclear matter} due to the important role of intermediary vector mesons in dilepton emission. 
In particular baryon-$\uprho$ interactions are known to strongly contribute to the observed melting of the $\uprho$ meson in the fireball formed in heavy-ion collisions from SIS18 to LHC energies~\cite{Hades19_AuAu}. 
A detailed understanding of these effects requires microscopic calculations that depend on these couplings.
%}

\begin{figure}[tbh]
  \begin{center}
      \includegraphics[width=1\textwidth]{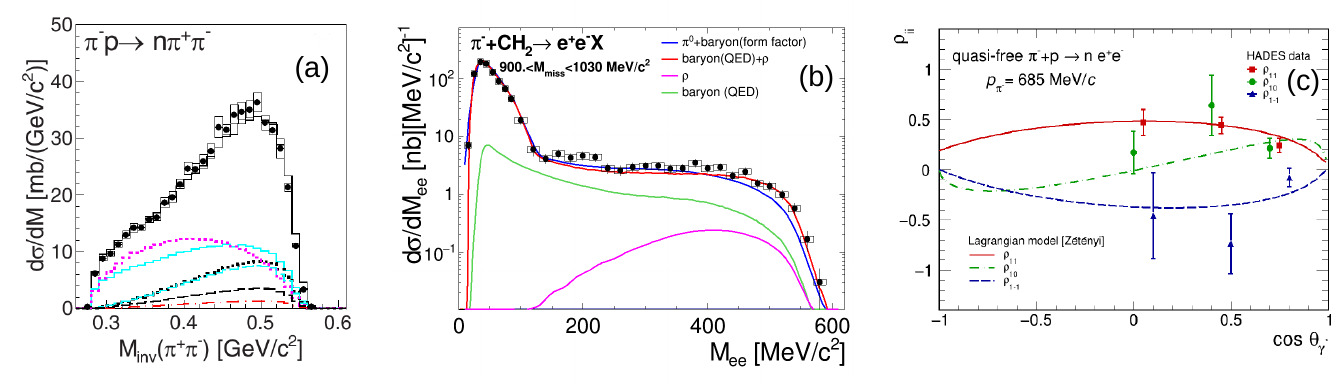}
   \vspace*{-7mm}
     \scriptsize
      \caption{\scriptsize Panel (a)
presents the  $\pi^+ \pi^-$ invariant mass distribution (black circles) measured with HADES in the \pim p$\to \pi^+\pi^-$ p reaction  for a pion incident momentum of 0.685~GeV/$c$ \cite{HadesPionbeamTwopi} compared to the Bonn-Gatchina PWA solution (black solid line histogram), with the subdivision into  the isobar $\Delta\pi$ (cyan), N$\sigma$ (dashed violet)  and  N$\uprho$ (violet) \rev{channels}. 
The $\uprho$ contributions from s-channels (green curves) as well as from D13 (blue curve) and S11 (red curves) partial waves are also displayed. 
Middle and right panels present the analysis of the quasi-free \pim~+~C$_2$H$_4~\to$n\ee\  reaction at $p_{\pi}$=0.685 GeV ($\sqrt{s}$=1.487 GeV), after selection of events with 0.9 $<$ M$_{\rm miss} < $1.03 \gevcc . 
Panels (b) \cite{yassine2023} and (c) \cite{yassine2024} Distribution of \ee\ invariant mass (M$_{\rm ee}$) compared to two different phenomenological approaches: (i) baryon Dalitz decay with form factor from \cite{Ramalho17} (red) and (ii) point-like baryon Dalitz decay (green curve) + $\uprho$ (violet) (see text for details). 
In the first case, the contribution from \piz\ Dalitz decay has also been added. 
Panel (c): Spin density matrix elements $\uprho_{11}$, $\uprho_{10}$ and  $\uprho_{1-1}$ extracted from the angular distributions of events with M$_{\rm ee} > 300$ \mevcc , compared to the ones derived from the $\uprho \rightarrow$ \pip \pim\ PWA. \label{fig:PionBeam2014}} 
  \end{center}
\end{figure}

\rev{During the 2014 commissioning run of the pion-beam facility with HADES, we measured double-pion production using a polyethylene ((CH$_2$)$_n$) and a carbon target in the second resonance region ($\sqrt{s} \simeq 1.49$\gev} \cite{HadesPionbeamTwopi}\rev{).} 
The obtained two-pion (\pip \pim\ and \pim \piz ) data samples have been included in the multichannel Partial Wave Analysis (PWA) of Bonn-Gatchina together with other world data on single and double pion production in photon, pion and electron induced reactions, which allowed for the extraction of the various final ($\Delta \pi, \uprho \pi$, N$\sigma$ ) and initial ($\frac{1}{2}^+$, $\frac{1}{2}^-$, $\frac{3}{2}^+, \cdots$) resonant and non-resonant states. As the information on \pip \pim\ and \pim \piz\  channels from other experiments only exist as total cross sections,  the HADES data were essential to constrain the $\uprho$ production (Fig.~\ref{fig:PionBeam2014}a) and to extract the branching ratios of the N(1440), N(1520) and N(1535) baryon resonances to the $\uprho$N channel, a basic information which is presently absent in the Review of Particle Physics.
This information has then been used for the interpretation of the dielectron production \cite{yassine2023, yassine2024}, by means of the Vector Meson Dominance (VMD) model, as explained below.
 \begin{table}[t]
\begin{center}
\scriptsize 
\begin{tabular}{lcccccccc}
\hline \hline
    channel &   $\epsilon_{AR}$
    & $\sigma_H$ & $\sigma_C$ & $\sigma_{Ag}$& \text{BR} & C$_2$H$_4$ target  & C target & Ag target\\ 
 &    &  \text{(mb)} & \text{(mb)}& \text{(mb)} & & $\dot{N}_H/\dot{N_{tot}}$ (shift$^{-1}$)&  $\dot{N}_C$(shift$^{-1}$) & $\dot{N}_{Ag}$(shift$^{-1}$)\\
    \hline 
    $\pi^-\pi^+$ \text{n}   & 0.14 & 10 &16 & 63 & 1 &  3.4 $\times 10^6$ /6.1 $\times 10^6$ & 3.8 $\times 10^6$ & 2.0 $\times 10^6$\\  
    $\pi^-\pi^0$ \text{p} & 0.09 & 6.5& 10.4 & 41 & 1 &  1.4 $\times 10^6$/2.6 $\times 10^6$ &   1.6 $\times 10^6$ & 8.6 $\times 10^5$\\  
  $\pi^0\pi^0$ \text{n} & 0.01 & 2 & 3.2 & 13 & 1 &  4.9 $\times 10^4$/8.8 $\times 10^4$ &  5.4 $\times 10^4$  & 2.9$\times 10^4$\\  
 $\text{K}^0 \Lambda$   & 0.04 & 0.56 & 1.85 & 7.3& 0.35 &  $1.9 \times 10^4$/$5.0 \times 10^4$ &  $4.3 \times 10^4$ & $2.4 \times 10^4$\\  
  $\text{K}^0 \Sigma^0$  & 0.04 & 0.24 & 0.79 & 3.1 & 0.35 & $8 \times 10^3$/$2.2 \times 10^4$ & $1.9 \times 10^4$ & $1.0 \times 10^4$\\  
  $\text{K}^+ \Sigma^-$ & 0.13 & 0.23 & 0.76 & 3.0 & 1 & $7.2 \times 10^4$/$1.9 \times 10^5$ & $1.7 \times 10^5$ & $9.0 \times 10^4$ \\  
  $\upeta \text{n}$  & 0.01 & 1.2 & 3.96 & 15.6 & 0.39 &  $1.2\times 10^4$/$3.1 \times 10^4$ & $2.6 \times 10^4$ & $1.4 \times 10^4$ \\  
  $\upomega \text{n}$  & 0.015 & 1.5 & 4.95 & 19.5 &0.89 &  $4.9 \times 10^4$/$1.3 \times 10^5$ & $1.1 \times 10^5$ & $4.0 \times 10^4$ \\  
    $\uprho \to $\ee\     &   0.25 & 2.1& 6.93 & 90.3  & 6\,$10^{-5}$ & 78/204 &  176 & 95 \\
    $\upomega \to$\ee\   & 0.31 & 1.7 & 5.61 & 73.1& 7.4\,$10^{-5}$ &  84/222 & 190 & 104\\  
   \hline \hline
\end{tabular}
\end{center}
\caption{\scriptsize Inputs for calculation of count rates in the different channels of interest  for the polyethylene (CH$_{2}$), carbon (C) and silver (Ag) targets for $\sqrt{s}$=1.76 \gev: reduction factors  due to the combined acceptance and reconstruction efficiency ($\epsilon_{AR}$), cross sections for \pimp\ ($\sigma_{\rm H}$), \pimC\ ($\sigma_{\rm C}$) and \pimAg\ ($\sigma_{\rm Ag}$) reactions, branching ratios (BR). The three last columns indicate the rate of reconstructed events per shift.  For the polyethylene target, the first and second numbers \rev{correspond} to interactions with protons and to the  total, respectively.}
\label{tab:ratesPEC}
\end{table}
We take these results  as a proof of concept of the Partial Wave Analysis method for HADES pion beam data in the two-pion production channels, which can be easily \rev{extended to} other hadronic channels in the third resonance region.

At a center-of-mass energy $\sqrt{s}$ around 1.73~GeV, many hadronic channels are open ($e.g.$ $K^0\Lambda$, $\Sigma^0 K^0$,  $\Sigma^+K^-$, $\upeta \text{n}$, $\upomega \text{n}$ ,...) in addition to the two-pion production. 
These channels will help to constrain \rev{the couplings of baryon states in this region to strangeness-carrying mesons} ($\Delta(1620)$ 1/2$^-$, $\Delta(1700)$ 3/2$^-$, N(1650) 1/2$^-$, N(1675) 5/2$^-$, N(1680) 5/2$^+$, N(1710) 1/2$^+$, N(1720) 3/2$^+$,$\cdots$) which are very poorly known. 
Our goal will therefore be to provide \rev{high-statistics differential data} for hadronic channels, including those with neutral mesons.
Together with complementary photoproduction data taken in a similar center-of-mass energy at CBELSA combined with our close collaboration with corresponding colleagues both from experiment and theory, {\it e.g.} Bonn-Gatchina group, we \rev{expect to foster our understanding of the light baryon spectrum in this energy interval}.

\rev{The two-pion channels, with cross sections of the order of 6.5--10~mb, can be measured with high statistics, resulting in new determinations of the 2$\pi$N channels, especially the $\uprho$N final state, which has a strong impact on medium effects, as mentioned above.} 
Resonant final states of the type N(1440)$\uppi$, N(1520)$\uppi$, $\cdots$ also attract much interest since they are related to \rev{the internal excitation structure of baryonic resonances, including single- and two-oscillator configurations with different decay patterns~\cite{Seifen2025}}, which might be a dominant decay mode in the case of the unobserved (or ``missing'') resonances. 
The study of 2$\uppi$ production in photo and electro-production was also recently used to provide evidence for a new N'(1720) 3/2$^+$ resonance \cite{Mokeev20}, with a mass lower by 20 \mev\ and very different $\uprho$ and electromagnetic couplings with respect to the known N(1720) 3/2$^+$ resonance.
By providing complementary data from pion induced reactions, our experiment can bring a timely contribution to this hadron structure highlight. 
The  data base for the $\upeta\text{n}$, K$\Lambda$ and $K\Sigma$ channels also need to be improved, which can in particular be used to firmly establish the existence of other resonances. 

\subsubsection*{Expected results}
Although good precision differential spectra 
for hadronic channels in \pimp\ reactions in the region between 1.68 and 1.8 \gev\ are badly missing, the total cross sections  are in general known with a precision of 10-20$\%$ and the reconstruction efficiencies of exclusive channels can be estimated based on previous analyses of HADES data, as in Ref.~\cite{HadesPionbeamTwopi} for the charged double pion production (see Table \ref{tab:ratesPEC}).

\rev{At these energies, strangeness conservation requires hyperons to be produced in association with a strange meson, resulting in exclusive channels such as K$^0\Lambda$, K$^0\Sigma^0$, and K$^+\Sigma^-$. For the K$^0\Lambda$ and K$^0\Sigma^0$ channels, hyperon production can be most efficiently reconstructed using K$^0_S$ production, corresponding to half of the cross section in the respective exclusive K$^0$ production channel.} The reconstruction of K$^0_S$ via their decay into  $\uppi^+ \uppi^-$  decay is used, with a branching ratio of 69$\%$ and an reconstruction efficiency of $\epsilon=4\%$, as deduced from full-scale GEANT simulations and validated by previous K$^0_S$ reconstructions in HADES experiments. The missing mass resolution of $\sigma=11$~\mev /$c^2$, as   measured in previous experiments, is sufficient to resolve the $\Lambda$ and $\Sigma^0$ states. For the $\Sigma^-$K$^+$, an efficiency of 13$\%$  is estimated. For $\upeta$ and $\upomega$ production, we use  the decay into $\uppi^+\uppi^-\uppi^0$, which can be reconstructed with  efficiencies  of 1 $\%$ and 1.5$\%$ for $\upeta$ and $\upomega$, respectively, as deduced using simulations that included the ECAL.

The count rates for the carbon target need also to be estimated, as they contribute to the statistical errors for the measurement of the \pimp\ reaction obtained by subtraction of pion-carbon interactions from the polyethylene data. Hadronic channels in $\pi$-A reactions are also interesting by themselves, for cold matter studies, as developed further below. We therefore give count rates for measurements on the carbon and silver targets for the above-mentioned channels. Inclusive meson production will be measured with even higher yields.
Reaction cross-sections measured for pion-nuclei reactions scale as $A^{2/3}$\cite{Allardyce73}. However, for the production of pions, which undergo strong absorption already in carbon, we used a ratio $\sigma_{\rm C}/\sigma_{p\rm } = 1.6$ as measured at $\sqrt{s}$=1.49~GeV. 
For the exclusive strangeness production channels,  we neglect the absorption and deduce the cross sections for the reaction on carbon and using the relation $\sigma \sim Z^{2/3}$, as mainly protons are involved. \\
\\
\\
\subsection{Electromagnetic couplings}

\subsubsection*{Motivation}
\rev{Virtual photons originating from baryonic matter provide a means to probe the electromagnetic currents and structure of the emitting hadrons.} From a heavy-ion perspective, virtual photons are not affected by final-state interactions, enabling direct exploration of the medium. \rev{From the standpoint of hadron physics, virtual photons provide direct access to the electromagnetic structure of hadrons and to electromagnetic transitions between baryon states, such as $N^*\rightarrow N\gamma^*$.} The HADES collaboration brings extensive expertise in utilizing virtual photons, primarily through dilepton detection, to study baryonic matter in heavy-ion collisions as well as in elementary processes.  

HADES experiments performed in the last years in proton-nucleus \cite{Agakishiev12_pNb} or nucleus-nucleus \cite{Hades19_AuAu} reactions have demonstrated an excess radiation of \ee\ above conventional sources, for invariant masses below the vector meson poles, attributed to emission out of the hot and dense stage of the reaction (fireball). 
\rev{The currently favored interpretation is that this radiation emerges} from decays of far off-shell $\uprho$ mesons with a strongly modified spectral function due to their coupling to baryonic resonances in hadronic matter.  

\rev{The study of elementary interactions, such as} \pp ,  quasi-free \np\ and  \pimp ~\cite{HADES_pp35_exclusive,Hades17_pnepem,Hades17_DeltaDalitz, yassine2023,yassine2024}, emphasized further the importance of intermediary $\uprho$ mesons in baryon resonance Dalitz decays (N/$\Delta \to$N\ee , in accordance with the VMD (see Fig.~\ref{fig:VMD},Right (b)).
Experiments using pion beams provide an increased sensitivity to these effects, as the resonances are excited in the $s$-channel in a narrow mass bin, corresponding to the pion beam momentum dispersion. 
This limits the \rev{overlap of many baryon states}, which is unavoidable in the case of nucleon-nucleon reactions. 
In addition, the \rev{identification of the exclusive channel} \pimp $\to$n \ee\ can be achieved by the simple detection of the \ee\ pair with a neutron missing-mass cut. 
Therefore, the \pimp $\to$n \ee\ reaction is an ideal tool to study the electromagnetic structure of baryon transitions in the region of small positive four-momentum transfer squared ($q^2 = M_{\rm ee}^2$), where vector meson poles play an important role.  
This information is complementary to the one obtained in electron scattering experiments in the space-like region ($q^2<0$) and is therefore needed for a global understanding of these transitions.
\par
\subsubsection*{Lessons from the 2014 data}
The differential cross sections in the \pimp $\to$n \ee\ reaction can be parameterized in a model independent way using different equivalent functions (helicity amplitudes, form factors or density matrix elements), which contain information on the  electromagnetic structure of the baryon transitions. 
In particular, in the context of the  HADES program,  the interest of studying the \pimp $\to$n \ee\ reaction is to check the validity of the VMD (see Fig.~\ref{fig:VMD}) for baryon electromagnetic transitions, which is used for the interpretation of \rev{\ee\ production in terms of the modified $\uprho$-meson spectral function}. 
The commissioning experiment performed in 2014 \rev{using polyethylene and carbon targets} confirmed the relevance of the \pimp $\to$n \ee\ channel to investigate time-like electromagnetic transitions and in particular test VMD.
%%%%%%%%%%%%%%JM
%
In the previous experiment in 2014, we measured the \ee\ production from polyethylene and carbon targets at a $\uppi$N center-of-mass energy of \sqrts=1.49 \gev , close to the pole of the N(1520) resonance. 
 \rev{The statistics for \ee\ events recorded on the carbon target were insufficient for an accurate subtraction to isolate \pimp\ interactions in the polyethylene data. However,} the quasi-free character of the \ee\ production in the \pimC\ reaction could be carefully checked and was taken into account in the modeling of the \pimP\ reaction using a quasi-free participant-spectator model.    The expected \mee\ invariant mass distributions  for the \pimp $\to$ n \ee\ channel can be estimated from  the known $\gamma n\leftrightarrow$ \pimp\ cross-sections in a simple QED calculation, assuming a point-like vertex  (dashed green line in Fig.
~\ref{fig:PionBeam2014}b).  To estimate the effect of time-like electromagnetic form factors,  which are expected to modify the yield at large M$_{\rm ee}$, we used models  recently developed  for the N-N(1520)~\cite{Ramalho17} and  N-N(1535) \cite{Ramalho20} transitions, where baryons are described with a quark core and a meson cloud, with parameters fitted to existing data in the space-like region. \rev{The time-like electromagnetic form factors cause a significant enhancement of the dielectron yield at large M$_{ee}$ w.r.t. the QED calculations} and provide a quantitative description of HADES data.  \par
We also adopted an alternative approach using  the $\uprho$ meson mass distribution obtained  in the PWA of the \pimp $\to$ \pip \pim\ n reaction, as explained above. The corresponding  \ee\ invariant mass distribution was calculated using VMD, which yields a dependence of the $\uprho \to $ \ee\ meson branching ratio $BR(M) \sim$ M$_{\rm ee}^3$ and this contribution was added to the point-like \ee\ emission. This "two-component" approach provides a modeling of the time-like electromagnetic baryon transition form factor in terms of two components: a photon and a $\uprho$ meson coupling. It is noteworthy that both approaches quantitatively yield the same results, in good agreement with the data.  
\subsubsection*{Extracting spin-density matrix elements}
HADES results also demonstrate that isoscalar ($\upomega$) contributions play a minor role in \ee\ production in the second resonance region, in contrast to the earlier predictions \cite{Lutz03}.  \par
 Further information, in particular on the transverse and longitudinal virtual photon polarization can be obtained from the angular distributions.  
 The density matrix formalism \cite{Speranza17} provides a convenient parameterization of the amplitudes $\left | A \right |$  at a given value of M$_{\rm ee}$ and emission angle of the virtual photon:

 \begin{align}
\left| A
\right|^2 \propto  &~ 4 k^2 [2\uprho_{00}(1-\cos^2\theta)+ 2\uprho_{11}(1+\cos^2\theta) \nonumber \\
   & + 2\sqrt{2}\sin(2\theta)\cos\phi Re\uprho_{10}
	+2\sin^2\theta Re\uprho_{1-1}\cos(2\phi)].
\end{align}

Here, $k^2$, $\theta$ and $\phi$ denote the momentum, \rev{polar and azimuthal angles of one of the leptons in the virtual-photon reference frame}, respectively, and $\uprho_{00}$, $\uprho_{11}$, $\uprho_{1-1}$ are the three independent  density matrix coefficients which can be expressed, for a transition of given spin and parity, as a function of helicity amplitudes or transition form factors.  
A method to extract spin density matrix elements from a fit to experimental data has been developed taking into account acceptance and efficiency effects \cite{yassine2024}. 
Despite the low statistics which affected the precision of the data in the 2014 experiment, and the low acceptance for backward virtual photon angles in the center-of-mass system, the coefficients were studied in three bins of center-of-mass angle of the virtual photon for $M_{e+e-}>0.14$ GeV/c$^2$, as shown in Fig.~\ref{fig:PionBeam2014}c.   
The deviations of  $\uprho_{11}$  from 0.5 and  $\uprho_{10}$  from 0  clearly demonstrates the contributions of virtual photons with longitudinal polarization, in contrast to real photons. 
Additionally, non-zero $\uprho_{-1,1}$ values indicate important contributions from transitions with spin larger than $\frac{1}{2}$, which is consistent with the dominance of the N(1520) resonance found in the two-pion analysis \cite{HadesPionbeamTwopi}. 
Indeed, the values of the spin density matrix elements are found in good agreement  with the predictions of the model of \cite{Speranza17} for the N-N(1520) transition, calculated with a VMD form factor. \par %
The density matrix formalism was  also  applied to analyze the angular distributions of pions from the $\uprho \to$\pip \pim\ component in the PWA solution, and  a good consistency was obtained with the coefficients   obtained from the dilepton distributions (curves on Fig.~\Ref{fig:PionBeam2014}). 
This is an additional, very direct test of the VMD approach, complementing the successful description of the invariant mass spectra. \par
\subsubsection*{Towards the third resonance regime}
The 2014 data yielded the  very first and promising information about the baryon electromagnetic transitions in the time-like region, motivating the exploration of the third resonance region, using the same approach. 
To demonstrate the feasibility of the measurement at higher energy, we developed full scale simulations of the $\uppi^-$ p $\to$ X\ee\ reaction. 
For the Dalitz decays of $\uppi^0$ and $\upeta$ mesons ($\uppi^0,\upeta \to \upgamma \text{e}^+ \text{e}^-$), which constitute the dominant contributions, the production cross sections from previous experiments  were used \cite{Baldini88,Shklyar13}. 
\ee\ emission from intermediate $\Delta(1232)$ Dalitz decay ($\Delta(1232)\to N \text{e}^+ \text{e}^-$)) is also taken into account, using  $\Delta\uppi$ cross sections from  a Bonn-Gatchina PWA solution of the $\uppi^-$p reaction, but has a much smaller contribution. 
For the simulation of the $\uppi^-\text{p}\to \text{n}~ \text{e}^+ \text{e}^-$ exclusive channels considered here as our signal, we used the two-component approach, which was successful for the data in the second resonance region, as described above.
Following the analysis of the $\uppi^-\text{p}\to \text{n}\upgamma$ reaction around \sqrts\ = 1.76 GeV, the N(1675) and $\Delta(1700)$ baryon resonances were identified as providing the most important  contributions, due to their large $\gamma$N couplings.  The $\uprho$ production cross sections is taken from the  HSD model \cite{Effenberger99inputs}  (2.1
~mb at \sqrts\ = 1.76 GeV), which accounts for the coupling of the $\uprho$ meson to baryon resonances.
\begin{figure}[tbh]
  \begin{center}
    \includegraphics[width=0.98\textwidth]{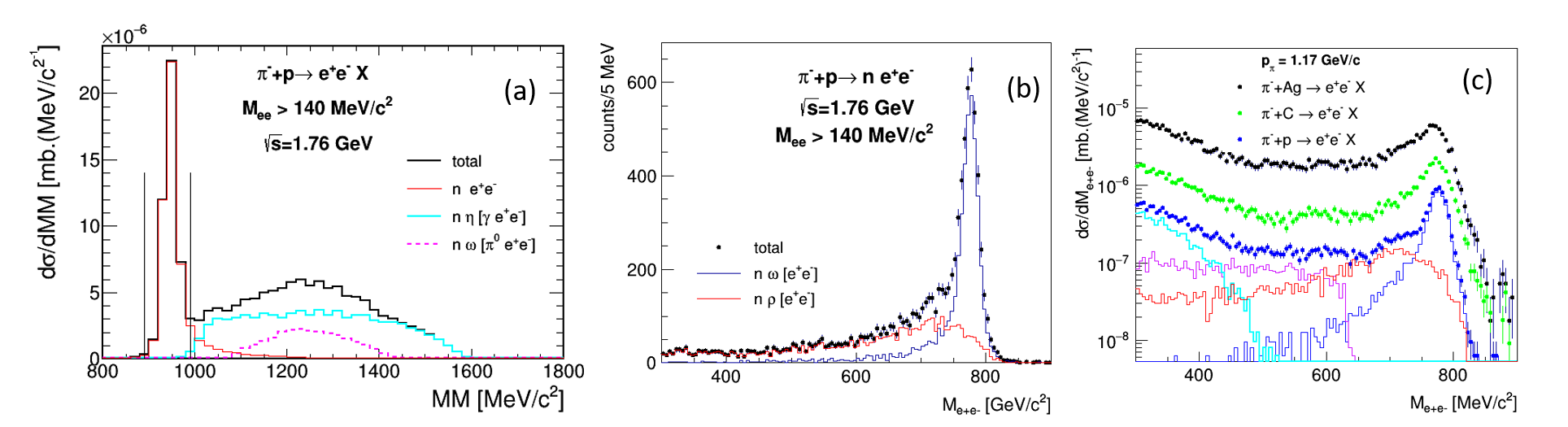}
  \end{center}
  \vspace*{-0.5cm}
  \caption{\scriptsize Simulations of the reaction $\uppi^-\text{p}\to \text{X}\text{e}^+ \text{e}^-$at p$_\uppi = 1.16$ \gevc\ including all known sources of dileptons. 
  (a) Missing mass (M$_\text{miss}$) distribution for events with invariant mass \mee > 140 \mevcc . 
  (b)  Expected distribution of counts in 5 \mev\ bins as a function of the invariant mass  after selection of the exclusive channel using the condition on the missing mass  $0.9 < \textrm{M}_\text{miss} < 1.03 $ \gevcc . 
  (c) Expected reconstructed inclusive \ee\ invariant mass spectra  for measurements on the silver target (black dots),  carbon target (green dots) and for interaction with protons in the polyethylene targets (blue dots). 
  The cyan, violet, red and blue curves display the $\upeta \rightarrow$\ee ,  $\upomega \rightarrow $\piz \ee , $\uprho \rightarrow$\ee  and $\upomega \rightarrow$\ee contributions.}
  \label{fig:MM_Minv}
\end{figure}
\rev{The resulting missing-mass and dielectron invariant-mass distributions are shown in Fig.~\ref{fig:MM_Minv}, demonstrating the expected separation of the exclusive channel and the relative contributions of the different dilepton sources.}
The higher (30 $\%$) branching ratio for the $\uprho$ decay into \ee\ of \rev{$6\times10^{-5}$}  w.r.t.\ the one \rev{at the pole} is due to the enhanced production of low mass $\uprho$ mesons due to \rev{their coupling to baryons}. Note, that these values strongly depend on the coupling of baryon resonances to the $\uprho$N channel, which will be extracted from the PWA analysis of the $\uppi^+\uppi^- n$ and $\uppi^0\uppi^- \text{p}$ \rev{final states}.

As the $\upomega$ spectral function is expected to be less sensitive to the coupling to baryons, we  took measured  cross sections for the $\uppi^- \text{p} \to \upomega \text{n}$ reaction, known with a precision of about 10$\%$ from measurements at Rutherford Laboratory using neutron detection and missing mass~\cite{Karami79}. 
For the combined acceptance and reconstruction efficiency  of \ee\ pairs produced from $\uprho$ and $\upomega$, we find values of 25$\%$ and 31$\%$ respectively from GEANT simulations, taking into account in particular the improved efficiency of the \ee\ pair reconstruction provided by the upgraded RICH detector.  

 The choice of a center-of-mass energy 40 \mev\ above the $\upomega$ production threshold allows for a detailed study of the $\upomega$ meson spectral function, which will serve as a reference for studies of omega absorption in nuclei.\par
 In the region below the $\upomega$ peak, significant interferences between the  \rev{isoscalar ("$\upomega$-like") and isovector ("$\uprho$-like")} \ee\ productions might be present, as predicted by \cite{Titov01,Lutz03} and could be studied as well.
 We also plan to analyze the electron angular distributions and extract the spin density coefficients $\uprho_{11}$, $\uprho_{10}$ and $\uprho_{1-1}$ to get more specific information on the nature of the transitions as a function of the \mee\ invariant mass.   We will therefore measure  the spin density matrix elements in the three bins of invariant masses ([300-600], [600-700] and [700-900] \mevcc ) and the three bins of the cosine of the virtual photon angle ($[-0.02,0.03]$, [0.3,0.6] and [0.6,0.95]).  According to the results of the previous experiment and detailed simulation studies, \rev{such a binning in angle is well suited to characterize electromagnetic transitions of baryonic resonances to the nucleon, in particular for resonances with spin larger than 1/2.} However, we would like to improve the precision of each fit by increasing the statistics  by at least a factor 10, i.e. we aim at 5000 \ee\ events in each mass bin. 
 
 Although there are only few  models for electromagnetic baryon transitions in the third resonance region,  \rev{the new data will offer the opportunity to extend approaches} like those proposed in \cite{Ramalho17,Speranza17} for the second resonance region. Previous investigations into the electromagnetic structure of baryon transitions have sparked significant theoretical interest in computing electromagnetic transition form factors. A range of models exists, utilizing non-perturbative techniques~\cite{ramalho24,zetenyi21}, as well as model-independent frameworks with methodologies allowing systematic improvements~\cite{alvaro23, an24}. \rev{These advancements provide a robust foundation for high-statistics and high-precision data to be compared with cutting-edge theoretical approaches.}  

\subsection{Exotic mesons}

The spectrum of scalar mesons remains one of the most intriguing and yet under-explored areas in hadron spectroscopy. This field has gained particular significance with the growing interest in exotic mesons. Within this sector, one expects to find the lowest glueballs, four-quark systems, hybrid states, and bound states of other mesons. Notably, many of these states manifest as scalar mesons, including:

\begin{itemize}
\item f$_{0}$(500) ($\sigma$), often interpreted as a mix of two-quark and four-quark components with a small glueball admixture;
\item f$_{0}$(980) and its isospin-1 counterpart a$_{0}$(980), viewed as two-quark bound states of $K\bar{K}$;
\item f$_{0}$(1370), typically identified as a two-quark state;
\item f$_{0}$(1500), suggested to be a mixture of two-quark and glueball contributions;
\item f$_{0}$(1710), which is likely the lowest pure glueball state.
\end{itemize}
These interpretations remain largely speculative, albeit grounded in decades of theoretical research spanning over 30 years. \rev{The lack of new and precise experimental data} has left many of these hypotheses \rev{untested}.

\begin{figure}[h]
  \begin{center}
   \includegraphics[width=0.6\textwidth]{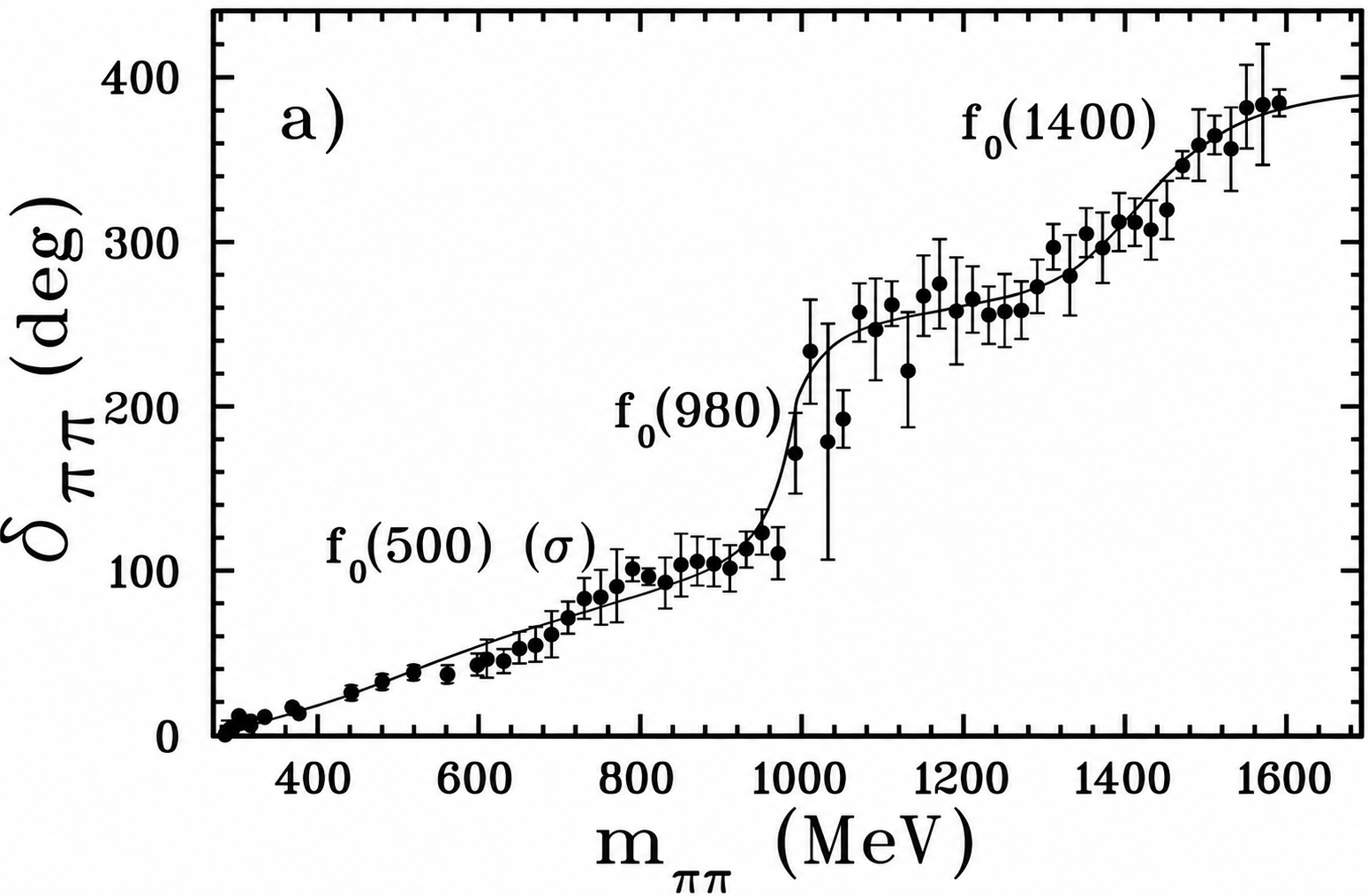}
         \end{center}
         \vspace*{-0.6cm}
          \caption{\scriptsize{The I=0, S-wave phase shift for the $\pi\pi$ interaction obtained from the fit to the experimental data \cite{Garcia11a}. The theoretical framework is based on dispersion relations with imposed crossing symmetry condition for the $\pi\pi$ interactions \cite{Kamin08, Kamin11, Garcia11,Garcia11a}.} } 
   \label{fig:mpipi}
  \end{figure}

The last comprehensive experimental investigations of scalar amplitudes were carried out by the CERN-Krakow-Munich collaboration in the late 1970s and early 1980s. These studies provided an extensive database and partial wave analyses of two-pion effective masses up to approximately \rev{1600 MeV} (see Fig.~\ref{fig:mpipi}). While these results were ground breaking for their time, they were also subject to significant uncertainties, such as the "up-down" ambiguity below 1000 MeV and multiple conflicting solutions.

All uncertainties and ambiguities were systematically addressed through theoretical developments. These efforts were grounded in Roy’s dispersion equations from the 1970s and extended by the GKPY dispersion equations introduced around 2010 \cite{Kamin08, Kamin11, Garcia11}. The new equations significantly reduced the average error in amplitude calculations, shrinking the allowable range of possible amplitudes by a factor of about six.
The work of the Madrid-Krakow group on GKPY equations, alongside efforts by the Bern group on Roy equations, culminated in major updates to the scalar meson sector of the Particle Data Tables in 2012. Notably, the meson f$_{0}$(500) received a new designation, along with substantially revised mass and decay width parameters.
Despite substantial theoretical advances in the analysis of amplitudes from the $\pi\pi$ threshold up to approximately 1800 MeV, the region around 1000 MeV remains particularly enigmatic. This range is dominated by the narrow f$_{0}$(980) state, which induces rapid amplitude variations. Analyzing these changes demands highly accurate, densely sampled experimental data (see Fig.~\ref{fig:mpipi}).
The internal structure of f$_{0}$(980) and its partner a$_{0}$(980) has been the subject of numerous theoretical studies, but none has provided a definitive characterization. Precise experimental data in the $\pi\pi$ and $K\bar{K}$ channels around the 1000 MeV region could significantly advance our understanding of meson spectroscopy and offer the first conclusive evidence for hypothesized mixed structures.
Using a pion beam at momenta around 2.3 GeV/$c$, HADES can deliver precise data in key channels essential for theoretical coupled-channel studies. These include the $\pi\pi$ and $K\bar{K}$ channels in the S0 wave, $\eta\pi$ and $K\bar{K}$ in the S1 wave, and $\omega\pi$ in the P1 wave.

\subsection{Rare eta meson decays}

One of the key challenges in modern experimental physics is to test the validity of Standard Model (SM) predictions and search for hints of new phenomena that fall outside its established framework. While the Standard Model provides an excellent description of particles and their interactions, it fails to explain certain fundamental issues, such as the matter-antimatter asymmetry in the Universe and the origin of neutrino masses.

An important area of both theoretical and experimental investigation is the CP problem inherent in the Standard Model. CP-violating processes have been observed in weak decays, such as $K_L$ leptonic decays (studied by the KTeV~\cite{cp1} and NA48~\cite{cp2} collaborations) and in $B$ meson decays (investigated by Belle~\cite{cp3} and BaBar~\cite{cp4, cp5, cp6, cp7}). However, the CP violation observed so far in the electroweak sector, originating from quark mixing described by the CKM matrix, is far too small to account for the observed matter-antimatter asymmetry.

Moreover, a possible violation of CP in the strong interaction would lead to a finite neutron electric dipole moment, which experimentally is considered to be very small, \rev{less than $1\times10^{-26}\,e\,\mathrm{cm}$}~\cite{abel}. The absence of CP violation in the strong interaction could potentially be explained by the existence of hypothetical light-mass QCD axions, which are not included in the Standard Model. \rev{This mechanism was proposed by Weinberg and Wilczek, with the axion arising as part of the electroweak Higgs sector and from a common breaking mechanism of the electroweak and Peccei--Quinn symmetries, with a predicted mass range of $\mathcal{O}$(100~keV $-$ 1~MeV).} Recent theoretical work~\cite{alv, alves2, sergi} has extended this hypothesis, suggesting that QCD axions with masses in the MeV range could exist. These axions are assumed to couple predominantly to the first-generation Standard Model quarks, be short-lived, decay primarily into e$^{+}$e$^{-}$ pairs, and have suppressed isovector coupling to cancel leading-order $\chi$PT contributions to axion-pion mixing. Under this framework, hadronic decays of $\eta$ and $\eta'$ mesons become valuable probes for QCD axions and axion-like particles. The most promising channels are three-body final states such as $\eta^{(\prime)} \to \pi^{0}\pi^{0}a (e^{+}e^{-})$ and $\eta^{(\prime)} \to \pi^{+}\pi^{-}a (e^{+}e^{-})$, where $a$ denotes the axion-like particle.
Additionally, the observation of a "bump-like" excess (the so-called Atomki Anomaly) in the invariant mass distribution of $e^{+}e^{-}$ pairs emitted during the de-excitation of specific states of $^{8}$Be and $^{4}$He nuclei~\cite{krasz16, krasz} has been interpreted as evidence of a piophobic QCD axion with a mass of 17 MeV, referred to as the $X(17)$ particle.

The HADES detector with its capability for detecting low-mass $e^{+}e^{-}$ pairs can be used for studies of the rare $\eta$ decays and particularly to search for the $X(17)$ particle.   
In this context, assuming a dominant decay of the potential QCD axion into a dilepton pair $e^{+}e^{-}$, one can benefit from \rev{the high reconstruction efficiency for lepton pairs with low invariant mass and low lepton momenta in HADES}.
Sensitivity to the coupling of the axial-vector axion to the $e^{+}e^{-}$ has been proven using the p+p data collected with HADES at 4.5 GeV \cite{x17}. 
The $\eta$ signal has been reconstructed  using the decay $\eta\to\pi^{+}\pi^{-}e^{+}e^{-}$ and the preliminary upper limit of the branching ratio BR($\eta\to\pi^{+}\pi^{-}$X17) has been estimated to be $<2.58\times10^{-5}$, however the multi-pion background contribution is very large resulting in a low signal to background ratio. The experiment with the pion beam offers a unique possibility to study the $\eta$ decay in a practically background-free exclusive reaction $\pi^{-} + p \to n + \eta$,  with a neutron tagging using the missing mass method. 

\clearpage
\section{{\it D:} Effective interactions}
\label{sec:effinteractions}

\subsection{Introduction}

\rev{This section focuses on experiments with pion beams on proton and nuclear targets aimed at extracting information on hyperon--nucleon and hyperon--meson interactions at low energies within the SU(3)-flavor framework.}

\rev{For the hyperon--nucleon interaction, the experiments address hyperon polarization in $\pi N$ interactions.} This information is essential for follow-up investigations into the \rev{spin degrees of freedom of hyperon--nucleon interactions}, utilizing secondary polarized hyperon "beams" in planned experiments at J-PARC. Additionally, it provides critical observables for partial-wave analyses in baryon spectroscopy studies, as outlined in Sec.~\ref{sec:structure}. Furthermore, pion beams interacting with nuclear targets offer an excellent source for hypernuclei production, whose properties provide deeper insights into the low-energy aspects of the hyperon-nucleon interaction.

In addition to studies on the hyperon-nucleon interaction, we also plan to investigate the hyperon-meson interaction through final-state interaction studies. These measurements complement CP-violation studies in the hyperon sector conducted by BESIII and HyperCP, particularly by providing precise determinations of the strong-phase difference via $\pi \Lambda$ final-state interactions.

\subsection{Hyperon polarization}

\rev{In the 1970s, a large transverse polarization of $\Lambda$ hyperons was discovered in unpolarized proton--beryllium collisions. Since then, similar polarization phenomena have been observed in deep-inelastic lepton--hadron scattering, hadron--hadron and hadron--nucleus reactions, as well as in electron--positron collisions. There exist many theoretical approaches, some of which are successful} \cite{DeGrand85}\rev{; however, none describes the world data consistently.}
\rev{Nonzero recoil polarization has also been observed for other hyperons, such as $\Sigma$ and $\Xi$.} 
Studying this phenomenon is crucial for understanding the mechanism of strangeness production, as well as for obtaining
 fundamental information about the hyperon-nucleon (Y-N) interaction and nuclear forces. This can be achieved by performing secondary two-body Y-N scattering experiments. Studies of polarization observables like analyzing powers or depolarization can give new insight in the nuclear forces and dynamics in \rev{hyperon--nucleon systems}.
\rev{Existing $\Lambda$ and $\Sigma^{0}$ polarization data from pion-beam experiments} \cite{Baker78, Crolius67, Penner02} \rev{suffer from large statistical and uncontrolled systematic uncertainties.}
 Precise polarization determination and identification of  angular regions with the highest polarization can be very useful for designing future measurements with secondary hyperon beams, like the J-PARC experiment
\cite{JPARC,JPARC_P86}.
 J-PARC will use high intensity $\pi^{+/-}$ beams to perform precise  \rev{N$^{\ast}$ baryon spectroscopy} and measurements of  spin observables of the $\Lambda$-p scattering with a polarized $\Lambda$ secondary beam \cite{JPARC_P86}.

Precise polarization determination along with differential cross sections are also important for \rev{partial-wave analyses} \cite{Thiel22} and for coupled-channel models \cite{Penner02} to identify various resonance contributions. \rev{For example, a measurement of the $\Sigma^{0}$ recoil polarization in $\pi^{-}p\to K^{0}\Sigma^{0}$ is crucial}  for the determination of the D$_{33}$(1700) and P$_{33}$(1920) resonances. \\
\indent \rev{Measurements of pion--proton interactions open the possibility to study the $\Lambda$ and $\Sigma^{0}$ self-polarization.}  The $\Lambda$ polarization is expected to be very large $\sim$100\% in this energy range  \cite{Baker78}, while for $\Sigma^{0}$ \cite{Crolius67} only few data points exist  with very large statistical errors.
To measure the Lambda polarization K$^{0}$ reconstruction via 2 pion decay ($\pi^{+}\pi^{-}$) is assumed and the third particle, either pion or proton, is needed to extract the polarization value $P_{\Lambda}$ using its angular distribution ($\cos(\beta)$) in the $\Lambda$ reference frame.  The angular distribution of the decayed pion or proton in the rest system of $\Lambda$ with respect to some reference axis is given by the formula:
\begin{equation}
    \frac{1}{N_{0}}\frac{dN}{d\cos(\beta)}=1+\alpha P_{\Lambda}\cos(\beta),
    \label{Eq:angDistrib}
\end{equation}
where N$_{0}$ is the "unpolarized" yield of the decay hadron, $\alpha$ is the decay asymmetry parameter, which was recently updated as $0.750\pm 0.009\pm0.004$ \cite{abli19}, $P_{\Lambda}= P_{\Lambda}(\cos\theta^{CM}
_{\Lambda})$ is the $\Lambda$ polarization and $\theta_{\Lambda}$ is the $\Lambda$ scattering angle.  \\
\begin{figure}[tbh]
  \begin{center}
   \includegraphics[width=0.45\textwidth]{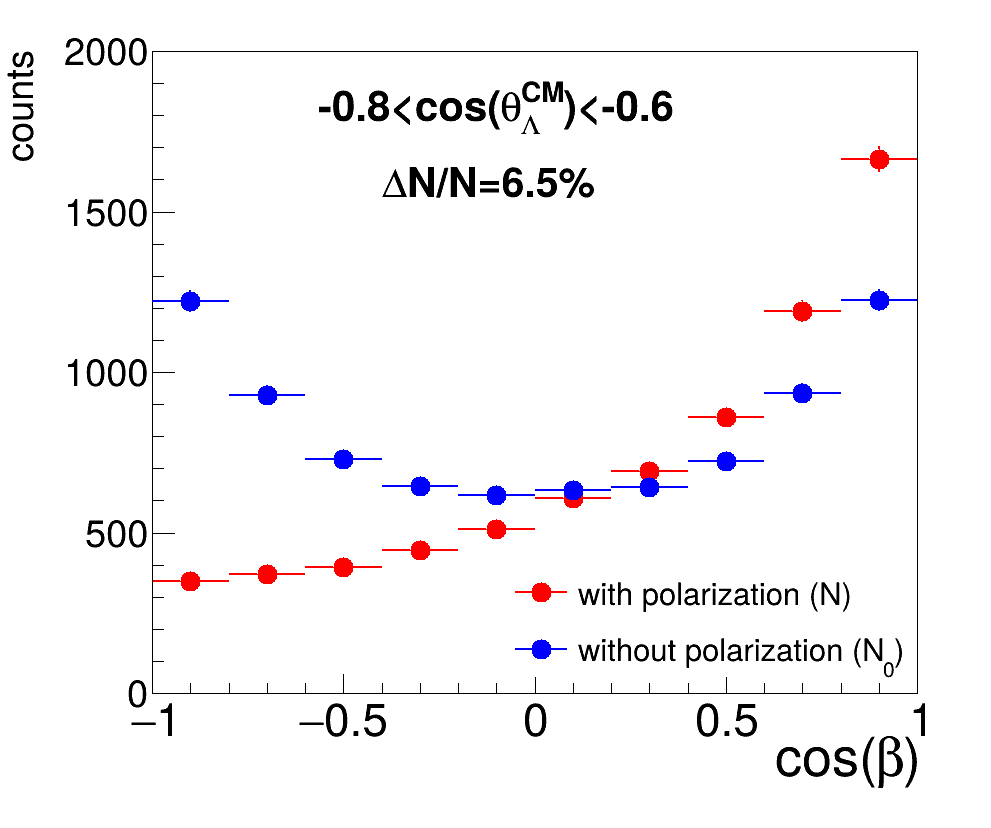}
          \includegraphics[width=0.45\textwidth]{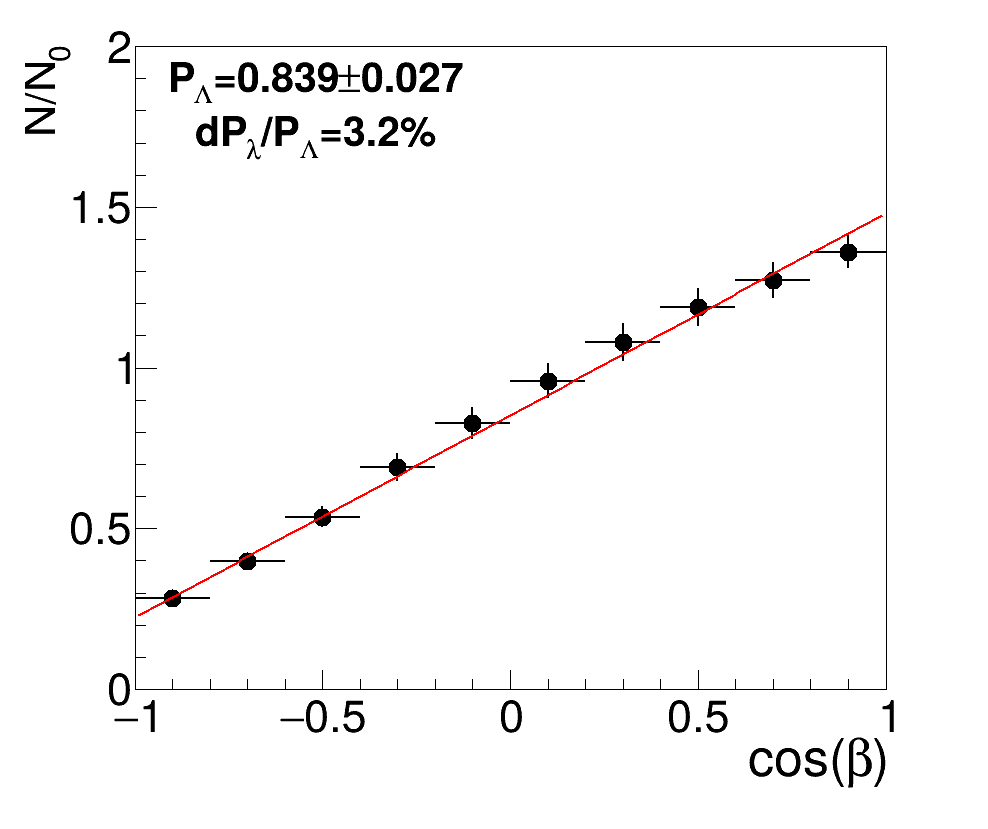}               \vspace*{-3mm}
        \caption{\scriptsize Left panel: expected statistics  in the  distribution of the emission angle of the \rev{decay proton} with (N - red points) and without polarization (N$_{0}$ - blue points) in a given range
of $\cos\theta_{\Lambda}^{CM}$ for $\pi^{-}p\to\Lambda K^{0}$ at $\sqrt{s}=1.73$ GeV for 7 shifts on the CH$_{2}$ target. The estimated statistical accuracy is $\sim6.5\%$. Right panel: same as in the left panel but
for the ratio of the "polarized" (N) and "unpolarized" (N$_{0}$) distributions. The estimated statistical accuracy of the \rev{polarization extracted from the linear fit is} $\sim3\%$.}
\label{fig:Polariz}
  \end{center}
\end{figure}

\indent To estimate the expected count rates, the full scale simulations of the $\pi^{-}p\to\Lambda K^{0}$ and $\pi^{-}p\to\Sigma^{0} K^{0}$  reactions have been performed including \rev{realistic angular distributions}
 of $\Lambda$\cite{Baker78,Penner02} and $\Sigma^{0}$ \cite{Crolius67}. The left panel in Fig.~\ref{fig:Polariz} presents a sample distribution of the expected "unpolarized" N$_{0}$ and "polarized" N (N$_{0}$ distribution
 weighted according to Eq.~\ref{Eq:angDistrib}) count rates \rev{as a function of $\cos(\beta)$} in a selected bin of  $\cos(\theta^{CM}_{\Lambda})$.
The polarization value has been obtained from a linear fit to the N/N$_{0}$ ratio distribution, as visible in Fig.~\ref{fig:Polariz}, right panel. We aim at a precise determination of the $\Lambda$ polarization at 5 energy points: $\sqrt{s}=1.67$ \gev ,  $\sqrt{s}=1.70$ \gev , $\sqrt{s}=1.73$ \gev, $\sqrt{s}=1.73$ \gev~ and  $\sqrt{s}=1.79$ \gev.\\
\indent The $\Sigma^{0}$ polarization will be measured in the longest run at $\sqrt{s}=1.76$ GeV with an expected statistical accuracy of 7\%.  Since $\Sigma^{0}$ decays electromagnetically into $\Lambda+\gamma$, the $\Lambda$ decay into $p\pi^{-}$ will be used as an analyzer of the $\Sigma^{0}$ self-polarization.\\
\indent So far, the \rev{$\Lambda$ and $\Sigma^{0}$ polarizations have been measured} with very low accuracy of $\sim20\%$\cite{Baker78,Crolius67}.
In this experiment we aim to measure the polarization with the unprecedent precision of $\sim6\%$. Such a precision enables stringent tests of microscopic models of hyperon production and spin transfer, and provides quantitative input to enable future experiments studying the spin and isospin dependence of the hyperon–nucleon interaction.\\

\subsection{Hypernuclei formation}
\begin{figure}[tbh]
  \begin{center}
    \includegraphics[width=0.42\textwidth]{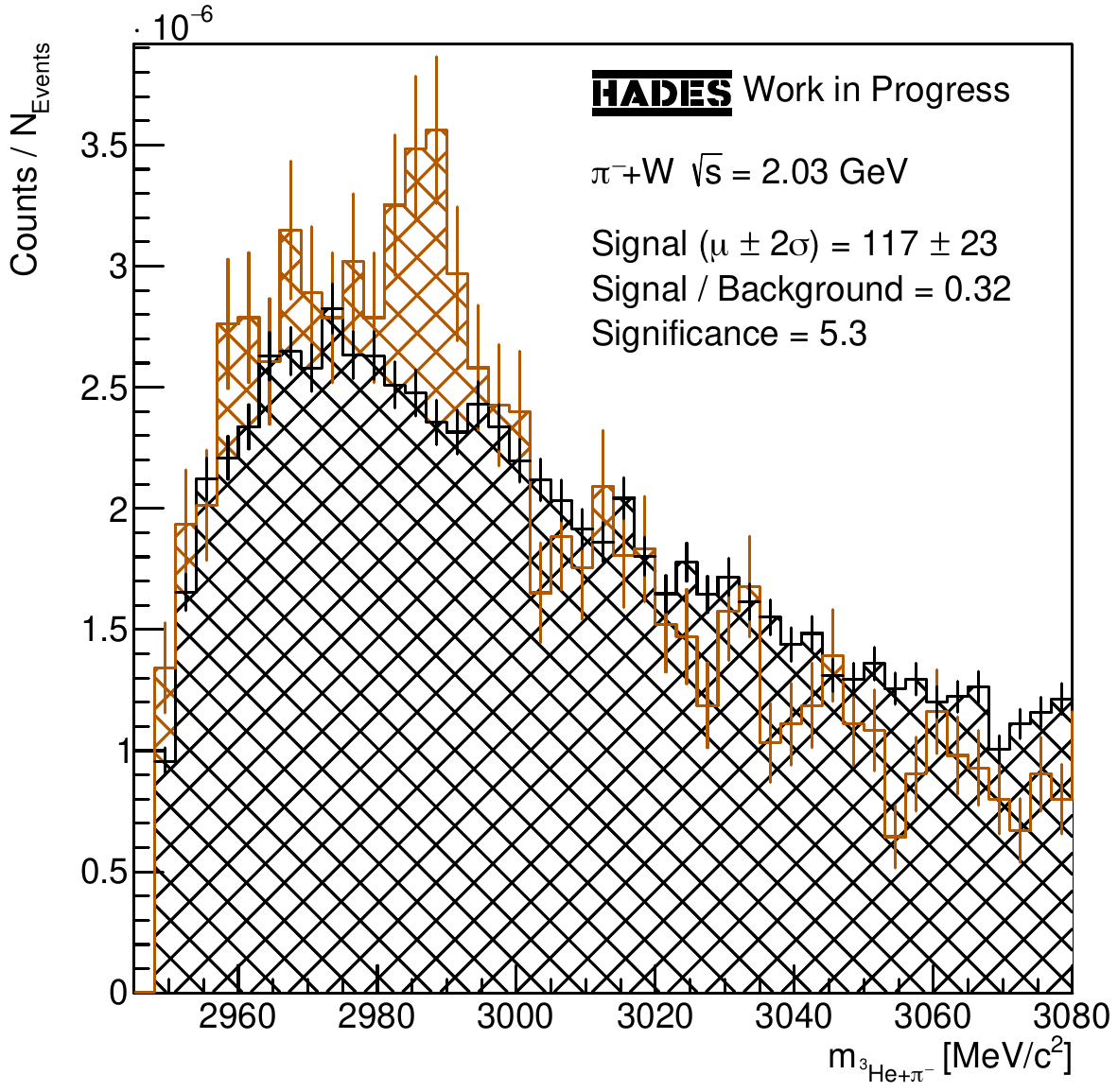} 
  \end{center}
   \scriptsize
   \vspace*{-0.5cm}
  \caption{\scriptsize Invariant mass distribution of $\pi^-$ and $^3\text{He}$ pairs from the $\pi^- + \text{W}$ data taken in 2014 (orange). A signal of the hypertriton with about 120 counts and a significance of 5.3\rev{$\sigma$} emerges above the mixed-event background (black).}
  \label{fig:Hyp}
\end{figure}

Hypernuclei offer an additional way to investigate the hyperon-nucleon ($Y$-$N$) interaction. The small recoil momentum of secondary particles produced in meson beams is particularly advantageous for the formation of hypernuclei~\cite{Kittiratpattana:2023atz}. 
Our research has demonstrated that the particle identification (PID) and vertex reconstruction capabilities of HADES, combined with the use of an artificial neural network (ANN), are well-suited for reconstructing hypernuclei and performing precise measurements of their lifetimes and branching ratios~\cite{SpiesPhD}. These capabilities allow the extraction of relevant information about the $Y$-$N$ interaction from hypernuclei data within the same experimental run. \\
Figure~\ref{fig:Hyp} shows the invariant mass distribution of $\pi^-$ and $^3\text{He}$ pairs from the $\pi^- + \text{W}$ data taken in 2014. A signal of the hypertriton with about 120 counts and a significance of 5.3$\sigma$ emerges above the mixed-event background.

The proposed measurements on cold nuclear matter aim to leverage significant improvements in the pion beam properties and detector technologies compared to the 2014 experiment. Hence, the anticipated future beam time will substantially improve upon the 2014 experiment, resulting in an overall gain factor of 100. This significant increase in efficiency and data collection capability translates into an expected production of \rev{approximately 10,000 hypernuclei}.
These yields represent an up to now unmatched amount, (almost an order of magnitude increase) of hypernuclei statistics at GSI and  will enable statistically robust studies of hypernuclei properties, including lifetimes, branching ratios, and formation dynamics.\\

\subsection{Hyperon-meson interaction}

The SM successfully describes fundamental particles in particle physics but does not provide a satisfactory explanation for the matter-antimatter asymmetry in the Universe. This phenomenon can be explained by the existence of processes that violate CP symmetry \cite{sak67}.  

\rev{The existence of CP violation consistent with SM expectations has been confirmed in kaon, beauty- and charm-meson decays, as well as in baryon decays.} A very sensitive test of CP violation and effects beyond SM is offered by the hyperons (Y) decays \cite{don85}.  The weak decays of hyperons involve two amplitudes: a parity conserving P-wave (parity-even), and a parity violating S-wave (parity-odd). CP violation can be observed if there is interference between CP-even and CP-odd terms in the decay amplitude. A CP violation signal in the hyperon decays has been extracted in the BESIII and  HyperCP \cite{hyperCP} experiments. The BESIII collaboration has developed a new and very sensitive method to study the CP violation effects using sequentially decaying entangled baryon-antibaryon pairs produced in the   $e^{+}e^{-}\to J/\psi\to \Lambda\bar{\Lambda}$/$\Xi^{-}\bar{\Xi^{+}}$ processes \cite{Perotti:2018wxm}. They have extracted \rev{several $\Xi^{-}$ and $\Lambda$ decay-asymmetry parameters, also for their charge-conjugate channels} \cite{abli19,abli22,abli24}. 
 
In this framework the leading-order contribution to the CP asymmetry A$^{Y}_{CP}$ is given as: A$^{Y}_{CP}\approx-$tan$(\delta_{P}-\delta_{S})$tan$(\xi_{P} - \xi_{S})$, where $\delta_{P}-\delta_{S}$ denotes the  strong-phase difference of the final-state interaction (FSI) 
between the $\Lambda$ and $\pi^{-}$ from the $\Xi^{-}$ decay,
and $\xi_{P}-\xi_{S}$ denotes the weak-phase difference. CP-violating effects would manifest themselves in a nonzero weak phase.
The weak-phase difference has been directly determined for the decay $\Xi^{-}\to\Lambda\pi^{-}$ using entangled $\Xi^{-}$ and $\bar{\Xi}^{+}$ \cite{abli22}. The strong-phase difference has been determined indirectly
using the two other decay parameters $\langle\phi^{\Xi}\rangle$ and $\langle\alpha^{\Xi}\rangle$. This is one of the most precise tests
of the CP symmetry for strange baryons. \par
 To gain direct and more precise information on the strong-phase difference and increase the sensitivity to CP violation effects, knowledge on the $\Lambda-\pi^{-}$ FSI, which enters the calculations, is needed. Such effect can be studied with the HADES detector and pion beams in the three-body $\pi^{-}+p\to\Lambda+\pi^{-}+K^{+}$ reaction.
Using a partial wave or Dalitz plot analysis, the $\Lambda-\pi^{-}$ FSI effects can be pinpointed. These studies will be complementary to measurements planned at the JLab facility with photon beams. 

\begin{figure}[tbh]
  \begin{center}
    \includegraphics[width=1.0\textwidth]{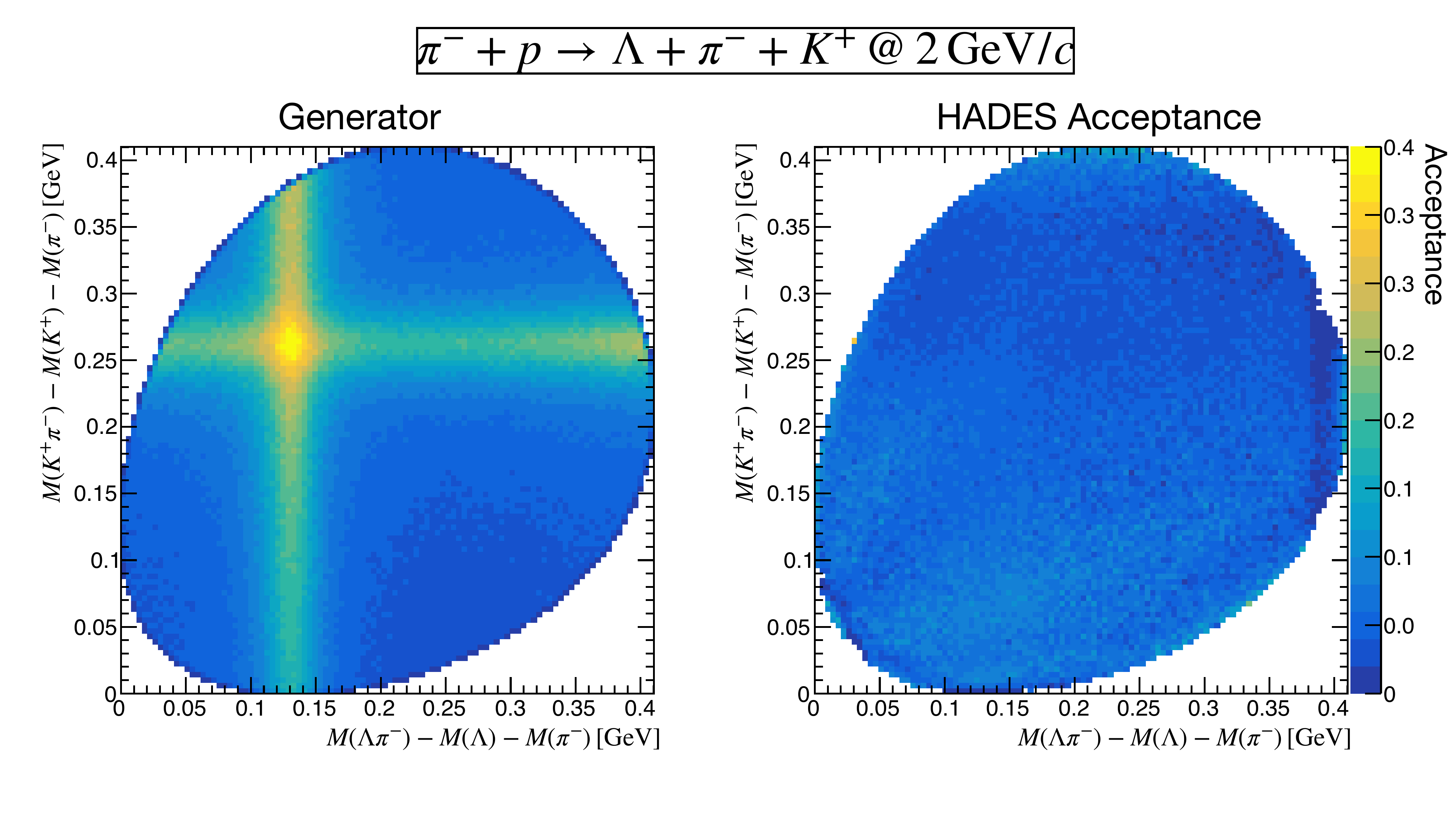} 
  \end{center}
   \scriptsize
   \vspace*{-1cm}
  \caption{\scriptsize Pluto simulation of the reaction $\pi^{-} + p \to \Lambda + \pi^{-} + K^{+}$ with $\Lambda\rightarrow \pi^- p$. {\it Left panel:} The output of the generator which includes \rev{an incoherent sum} of phase-space model plus a model incorporating the two intermediate $\Sigma(1385)$ and $K^*(892)$ resonances. {\it Right panel:} The acceptance of HADES for which all the final-state particles are registered by the HADES central detector or forward wall. Moreover, the momenta of all final-state particles are required to be larger than 100~MeV/$c$.}
  \label{fig:Lambda_pion_pluto}
\end{figure}

Figure~\ref{fig:Lambda_pion_pluto} illustrates the mass correlation between the $\pi K$ and $\pi \Lambda$ systems for the reaction $\pi^{-} + p \to \Lambda + \pi^{-} + K^{+}$ with $\Lambda\rightarrow p\pi^-$ at a pion-beam momentum of 2~GeV/$c$, based on Pluto simulations. The generator incorporates a mixture of a phase-space model and models with intermediate states involving the $\Sigma(1385)$ and $K^*(892)$. The left panel displays the generated spectrum, while the right panel shows the HADES acceptance, where all final-state particles are detected. \rev{It is noteworthy that the setup provides nearly complete phase-space coverage of the reaction with all final-state particles detected.}  
%}\\

\clearpage
%
%-------------------------------------------------------------
\section{Experimental set-up }
%-------------------------------------------------------------
%
\label{sec:exp-setup}
\subsection{The (upgraded) HADES facility}
HADES is a charged-particle detector consisting of a 6-coil toroidal magnet centered around the beam axis and six identical detection sections located between the coils, \rev{covering almost the full azimuthal angle and polar angles from $15^\circ$ to $85^\circ$}. Each sector is equipped with a Ring-Imaging Cherenkov (RICH) detector followed by Mini-Drift Chambers (MDCs), two in front of and two behind
the magnetic field, as well as a scintillator hodoscope
(TOF) and a Resistive Plate Chamber (RPC). At the end
of the system a forward hodoscope used for event plane
determination is located. The RICH detector is used
mainly for electron/positron identification, the MDCs are
the main tracking detectors, while the TOF and RPC are
used for time-of-flight measurements in combination with
a diamond start-detector located in front of the 15-fold
segmented target. The trigger is based on the hit multiplicity in the TOF covering a polar angle range between 45$^\circ$ and 85$^\circ$. A detailed description of the HADES detector is given in~\cite{Agakishiev:2009}.

The measurements we propose on baryonic resonances and cold hadronic matter will take advantage of foreseen improvements of the pion beam properties. In addition, the HADES set-up will include four new devices compared to the previous pion beam experiment in 2014:
\begin{youitemize}
\item The new RICH photon detector, successfully operated already in the latest HADES  experiments in 2019, 2022, 2024, and 2025 as part of the FAIR Phase-0 physics program, provides an increased electron reconstruction efficiency of about a factor of three~\cite{Becker:2023}.
 \item  The ECAL, also used in 2019, 2022, 2024, and 2025, extends the capacity of HADES and provides high precision differential cross sections for hadronic channels to exit channels containing neutral mesons and photons.
\item A new t0 and beam tracking detector based on Low Gain Avalanche Diode technology will be installed in front of the HADES target and will replace the Si tracker and diamond detector used in 2014. With three stations in front of the target, pion beam particles hitting the target can be selected and background rejected, hence providing a more accurate normalization of beam pions interacting with target nuclei. 
The trigger of the acquisition will be provided by a coincidence between a fast signal in the t0 detector and at least two signals in the RPCs or TOF-scintillators.
This will induce no bias, since all the channels we want to measure have two charged particles in the final state.
\item The new straw tube Forward Detector was already operated in 2022 and covers the angular range between 0.5 and 6$^{\circ}$, which can be useful in particular for elastic scattering measurements, which are useful for the normalization using the existing EPECUR data \cite{EPECUR} and for efficiency checks.
\end{youitemize} 

\subsection{The GSI pion-beam facility}

\begin{figure}[tbh]
  \begin{center}
    \includegraphics[width=0.95\textwidth]{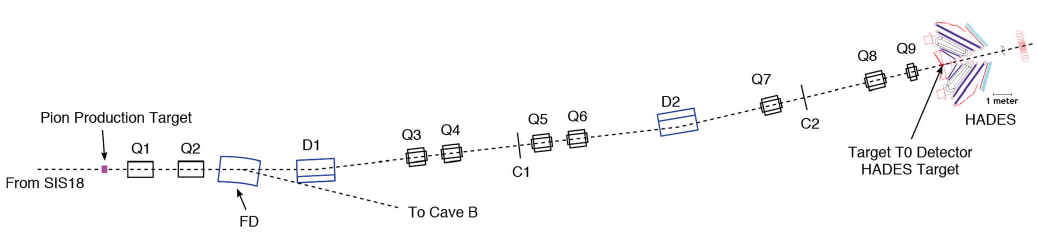} 
  \end{center}
   \scriptsize
   \vspace*{-0.5cm}
  \caption{\scriptsize Schematic overview of the beam line between the pion production target and the HADES cave. The dipole magnets (D), quadrupoles (Q), tracking detectors (C1, C2) and the target T0 detector are included in the transport simulations. The FD dipole used to direct the beam to cave B is treated as a passive element. Figure is taken from~\cite{Diaz02}.}
  \label{fig:piontracker}
\end{figure}

The GSI accelerator complex with a synchrotron of 18~Tm
maximum rigidity delivers protons and ions up to beam
energies of 4.5~GeV and 2~AGeV, respectively. A dedicated
target station for the production of secondary beams \rev{was already built in the 1990s} and serves several
caves. \rev{The HADES experiment is connected via a beam line with two tilted dipole magnets, D1 and D2, which elevate the standard beam tube height by 0.5~m up to the central axis of the spectrometer.} The beam line is depicted in
Fig.~\ref{fig:piontracker}. The HADES target point is located about 33.5~m
downstream of the production target. Position sensitive
beam tracking detectors C1 and C2 are mounted close
to the intermediate focal planes between the two dipoles
(C1) and between the quadrupoles inside the HADES cave
(C2). These quadrupoles (Q7, Q8, Q9 with $\ell$ = 1~m and
$\ell$ = 0.4~m) can be adjusted individually in position and are
relevant for the focusing conditions at the HADES target
point.

The secondary pion beam is generated using an extracted primary \( ^{14}\mathrm{N} \) beam at the highest possible intensity, impinging on a \( 10\,\mathrm{cm} \)-long beryllium target for all energies, except at the highest energy of \( \sqrt{s} = 2.35\,\mathrm{GeV} \). At this energy, a primary proton beam  impinging on beryllium is required to achieve sufficient intensity for the secondary pion beam.

The experiment in 2014 was realized using 36 shifts with an average of $2.2\cdot 10^5\,$ pions per second on the HADES target produced by a primary beam intensity of $4\cdot 10^{10}$  nitrogen ion/s.  
We used moderate extraction times to realize a 30\% duty cycle. 
\rev{The estimate of the achievable pion flux considers the following improvements with respect to the previous experiment.} 
First, the flux of $\pi^-$ with $p = 1.1$ \gevc\ at the production target is a factor of two higher compared to $\pi^-$ with momentum $p = 0.7$ \gevc, due to the forward  boost of the produced pions \cite{Diaz02}. Second, an additional increase by a factor two is expected using the maximum SIS18 primary beam intensity of $7\cdot 10^{10}$/s and taking into account improvements of the slow beam extraction, and the shielding added in the extraction area. 
The pions in a selected momentum range \rev{with a bandwidth of about 1.7\% (RMS)} are transported up to the HADES target by a set of nine quadrupoles and two dipoles. 
We will use five dipole settings to select different incident  pion beam momenta, as explained below. The quadrupole settings will be adjusted according to second order beam optics calculations to optimize the acceptance and focus requirements of the pion beam line.\par
We will use five different targets: a polyethylene (CH$_2$) block with a length of  $4.6\,$cm, a carbon (C) target consisting of 7 segments, each $3.6\,$mm long,  and interspersed by $7.1\,$mm,  a silver (Ag) target consisting of 15 segments, each $0.47\,$mm long and interspersed by about $3.6\,$mm and iron (Fe) and lead (Pb) targets each with thickness of 8 mm. All targets have the same 12\,mm diameter.
\rev{The choice of a polyethylene target is dictated by the larger density (factor 2) of protons ($3.8\cdot 10^{23}\,$cm$^{-2}$) compared to the available \lh\ target and by the absence of background from interactions with a target vessel.} The C and Ag target are segmented to reduce the energy loss due to bremsstrahlung and the background due to conversion of real photons, which impact the quality of \ee\ measurements.
The effective luminosities, calculated using a duty factor $\epsilon_{\rm duty}=0.5$ and a DAQ efficiency $\epsilon_{LT}=0.5$, amount for interactions with protons and carbon nuclei in the polyethylene target to $L$$^{\rm eff}_{\rm H}$ = 84.4  and 42.2~mb$^{-1}\cdot {\rm s}^{-1}$, respectively. For interactions in the carbon and silver targets the effective luminosities are $L$$^{\rm eff}_{\rm C}$= 58.4~mb$^{-1}\cdot {\rm s}^{-1}$ and $L$$^{\rm eff}_{\rm Ag}$= 8.05 ~mb$^{-1}\cdot {\rm s}^{-1}$, respectively.
\clearpage
\section{\rev{Summary and conclusion}}

\rev{This paper outlines a broad physics programme relevant for nuclear, hadron, heavy-ion, and astrophysics. In combination with the pion beam, HADES offers a unique experimental setting that connects precision studies of elementary $\pi N$ reactions with hadron propagation and interactions in cold nuclear matter. The programme addresses baryon spectroscopy and electromagnetic structure, in-medium hadron properties, strangeness and hypernuclear physics, rare decays, and precision input for neutrino--nucleus reaction modelling.}

\rev{The measurements will provide important benchmarks for microscopic and transport calculations and controlled reference data between elementary reactions and the hot and dense matter created in heavy-ion collisions. The pion-beam programme therefore represents an important part of the ongoing and future GSI/FAIR research programme and will extend our understanding of QCD in the strong-coupling regime.}

\newpage
\bibliographystyle{unsrt}
\bibliography{bibpionproposal}
\newpage
\newpage
\newpage

\section{The HADES Collaboration}
\label{sec:collaboration}

\begin{center}
%\begin{footnotesize}
\textrm{
R.~Abou~Yassine$^{7,14}$, J.~Adamczewski-Musch\orcidlink{0000-0002-7586-8504}$^{6}$, G.~Appagere\orcidlink{0009-0002-3969-6618}$^{16,*}$, M.~Becker$^{11}$,
A.~Blanco\orcidlink{0000-0001-9827-8294}$^{2}$, C.~Blume\orcidlink{0000-0002-6800-3465}$^{9,6,f}$, M.~Bohman\orcidlink{0009-0007-6118-8683}$^{17,*}$, Y.~Bondar\orcidlink{0000-0003-2773-9668}$^{4}$, L.~Chlad\orcidlink{0000-0003-3814-2920}$^{15,h}$,
I.~Ciepa{\l}\orcidlink{0000-0003-0936-3054}$^{4}$
% ...
}

%\end{footnotesize}
\end{center}
%%
%\institute{
\mbox{} \\[-8bp]
\mbox{$^{1}$Ruhr-Universit\"{a}t Bochum, 44801~Bochum, Germany}\\
\mbox{$^{2}$LIP-Laborat\'{o}rio de Instrumenta\c{c}\~{a}o e F\'{\i}sica Experimental de Part\'{\i}culas, 3004-516~Coimbra, Portugal}\\
\mbox{$^{3}$AGH University of Krakow, Faculty of Physics and Applied Computer Science, 30-059~Krakow, Poland}\\
\mbox{$^{4}$Institute of Nuclear Physics, Polish Academy of Sciences, 31342~Krak\'{o}w, Poland}\\
\mbox{$^{5}$Smoluchowski Institute of Physics, Jagiellonian University of Cracow, 30-059~Krak\'{o}w, Poland}\\
\mbox{$^{6}$GSI Helmholtzzentrum f\"{u}r Schwerionenforschung GmbH, 64291~Darmstadt, Germany}\\
\mbox{$^{7}$Institut f\"{u}r Kernphysik, Technische Universit\"{a}t Darmstadt, 64289~Darmstadt, Germany}\\
\mbox{$^{8}$Institut f\"{u}r Strahlenphysik, Helmholtz-Zentrum Dresden-Rossendorf, 01314~Dresden, Germany}\\
\mbox{$^{9}$Institut f\"{u}r Kernphysik, Goethe-Universit\"{a}t, 60438 ~Frankfurt, Germany}\\
\mbox{$^{10}$Physik Department E62, Technische Universit\"{a}t M\"{u}nchen, 85748~Garching, Germany}\\
\mbox{$^{11}$II.Physikalisches Institut, Justus Liebig Universit\"{a}t Giessen, 35392~Giessen, Germany}\\
\mbox{$^{12}$Department of Mechanical Engineering, Frederick University, 1036~Nicosia, Cyprus}\\
\mbox{$^{13}$Faculty of Science, Palack\'{y} University Olomouc, 779 00~Olomouc, Czech Republic}\\
\mbox{$^{14}$}\parbox[t]{0.9\textwidth}{Laboratoire de Physique des 2 infinis Irene Joliot-Curie, Universite Paris-Saclay, CNRS-IN2P3, F-91405~Orsay, France}\\
\mbox{$^{15}$Nuclear Physics Institute, The Czech Academy of Sciences, 25068~Rez, Czech Republic}\\
\mbox{$^{16}$Department of Physics, Stockholm University, ~Stockholm, Sweden}\\
\mbox{$^{17}$Institutionen for fysik och astronomi, Uppsala universitet, 75120~Uppsala, Sweden}\\
\mbox{$^{18}$Uniwersytet Warszawski, Instytut Fizyki Do\'{s}wiadczalnej, 02-093~Warszawa, Poland}\\
\mbox{$^{19}$Warsaw University of Technology, Faculty of Physics, 00-662~Warsaw, Poland}\\
\mbox{$^{20}$Bergische Universit\"{a}t Wuppertal, 42119~Wuppertal, Germany}\\
\\
\mbox{$^{*}$ members of the PANDA@HADES collaboration}\\
\mbox{$^{a}$ also at Helmholtz Research Academy Hesse for FAIR (HFHF), Campus Giessen, ~Giessen, Giessen}\\
\mbox{$^{b}$}\parbox[t]{0.9\textwidth}{also at Instituto Politecnico de Coimbra, Instituto Superior de Engenharia de Coimbra, 3030-199~Coimbra, Portugal}\\
\mbox{$^{c}$ also at Doctoral School of Exact and Natural Sciences, Jagiellonian University, ~Cracow, Poland}\\
\mbox{$^{d}$}\parbox[t]{0.9\textwidth}{also at Helmholtz Research Academy Hesse for FAIR (HFHF), Campus Darmstadt, 64390~Darmstadt, Germany}\\
\mbox{$^{e}$ also at Technische Universit\"{a}t Dresden, 01062~Dresden, Germany}\\
\mbox{$^{f}$}\parbox[t]{0.9\textwidth}{ also at Helmholtz Research Academy Hesse for FAIR (HFHF), Campus Frankfurt, 60438~Frankfurt am Main, Germany}\\
\mbox{$^{g}$ also at Charles University, Faculty of Mathematics and Physics, 12116~Prague, Czech Republic}\\
\mbox{$^{h}$ also at Czech Technical University in Prague, 16000~Prague, Czech Republic}\\
\mbox{ e-mail: hades-info@gsi.de (J.~Stroth)}\\
%}
\section*{Acknowledgments}
We gratefully acknowledge support by the following grants:
SIP JUC Cracow, Cracow (Poland), National Science Centre 2016/23/P/ST2/04066 POLONEZ, National Science Centre through grant SONATA-BIS no. 2023/50/E/ST2/00673, National Science Centre, Poland, 2017/26/M/ST2/00600; INP Cracow (Poland), National Science Centre grant nb. 2023/49/B/ST2/00652;WUT Warsaw (Poland) No: 2020/38/E/ST2/00019 (NCN), IDUB-POB-FWEiTE-3; TU Darmstadt, Darmstadt (Germany), VH-NG-823, DFG GRK 2128, DFG CRC-TR 211, BMBF:05P18RDFC1, HFHF (Campus Darmstadt), ELEMENTS 500/10.006, GSI F\&E, EMMI GSI Darmstadt; Goethe-University, Frankfurt (Germany), BMBF:05P12RFGHJ, GSI F\&E, HFHF (Campus Frankfurt), ELEMENTS 500/10.006; BU-Wuppertal, Wuppertal (Germany), BMBF 05P24PX1; JLU Giessen, Giessen (Germany), BMFTR 05P24RG6; IJCLab Orsay, Orsay (France), CNRS/IN2P3; NPI CAS, Rez, Rez (Czech Republic), MSMT LM2023060, MSMT OP JAK CZ.02.01.01/00/23$_{-}$015/0008181; The Swedish Research Council and the Knut and Alice Wallenberg foundation (Sweden). The authors thank the PANDA collaboration for the support with resources. This experiment was part of the FAIR-Phase0 program. The publication is funded by the OpenAccess Publishing Fund of GSI Helmholtzzentrum fuer Schwerionenforschung.
\end{document}